\documentclass[aps,prb,twocolumn,superscriptaddress,showpacs,floatfix,10pt]{revtex4-1} 
\usepackage{graphicx}

\usepackage[export]{adjustbox}

\usepackage{hhline}
\usepackage{multirow}

\usepackage{pifont}
\usepackage{bm} 

\usepackage{floatrow}
\usepackage[caption=false]{subfig}

\usepackage{xcolor}

\newcommand{\fref}[1]{Fig.~\ref{#1}}
\newcommand{\sref}[1]{Section~\ref{#1}}
\newcommand{\tref}[1]{Tab.~\ref{#1}}

\newcommand{\bra}[1]{\ensuremath{\langle #1|}}
\newcommand{\ket}[1]{\ensuremath{|#1\rangle}}

\newcommand{\com}[1]{}

\renewcommand\Im{\hbox{Im}}

\newcommand{\red}[1]{\textcolor[rgb]{0.89, 0.1, 0.11}{#1}}
\newcommand{\blue}[1]{\textcolor[rgb]{0.12, 0.47, 0.71}{#1}}

\newcommand{\green}[1]{\textcolor[rgb]{0.2, 0.63, 0.17}{#1}}
\newcommand{\yellow}[1]{\textcolor[rgb]{1., 0.5, 0.0}{#1}}

\newcommand{\svek}{%
        \mathbf}

\newcommand{\etal}{{\it et al.}}

\newcommand{\out}[1]{}

\renewcommand{\deg}{\ensuremath{^\circ}}

\def\XXint#1#2#3{{\setbox0=\hbox{$#1{#2#3}{\int}$}
\vcenter{\hbox{$#2#3$}}\kern-.5\wd0}}

\newcommand{\cbp}{Ce$_3$Bi$_4$Pt$_3$}
\newcommand{\csp}{Ce$_3$Sb$_4$Pt$_3$}

\newcommand{\cbpd}{Ce$_3$Bi$_4$Pd$_3$}
\newcommand{\cspd}{Ce$_3$Sb$_4$Pd$_3$}

\begin{document}

\title{Isoelectronic tuning of heavy fermion systems: \\
 Proposal to synthesize Ce$_3$Sb$_4$Pd$_3$} 

\author{Jan M. Tomczak}
\email{tomczak.jm@gmail.com}
\affiliation{Institute of Solid State Physics, Vienna University of Technology, A-1040 Vienna, Austria}

\begin{abstract}
The study of (quantum) phase transitions in heavy-fermion compounds relies on a detailed understanding of 
the microscopic control parameters that induce them. While the influence of external pressure is rather straight forward,
atomic substitutions are more involved. Nonetheless, replacing an elemental constituent of a compound with an isovalent atom
is---effects of disorder aside---often viewed as merely affecting the lattice constant.
Based on this picture of chemical pressure, the unit-cell volume is identified as an empirical proxy for the Kondo coupling.
Here instead, we propose an ``orbital scenario'' in which the coupling in complex systems can be tuned by isoelectronic substitutions
with little or no effect onto cohesive properties.
Starting with the Kondo insulator \cbp, we consider---within band-theory---isoelectronic substitutions of the pnictogen (Bi$\rightarrow$Sb) and/or the precious metal (Pt$\rightarrow$Pd).
We show for the {\it isovolume}  series Ce$_3$Bi$_4$(Pt$_{1-x}$Pd$_x$)$_3$ that the Kondo coupling is in fact substantially modified 
by the different {\it radial extent} of the $5d$ (Pt) and $4d$ (Pd) orbitals, while spin-orbit coupling mediated changes are minute.
Combining experimental Kondo temperatures with simulated hybridization functions, we also predict 
effective masses $m^*$, finding excellent agreement with many-body results for \cbp.
Our analysis motivates studying the so-far unknown Kondo insulator \cspd, for which we predict $m^*/m_{band}=\mathcal{O}(10)$.
\end{abstract}

\pacs{71.27.+a, 75.30.Mb, 71.30.+h}


\maketitle

\floatsetup[figure]{style=plain,subcapbesideposition=top}

\section{Introduction.}

The physics of materials with $f$-electrons is determined by a delicate interplay
of energy scales. The competition between the latter can be tuned by various means---including external fields, pressure, carrier doping---resulting in enormously rich phase diagrams%
\cite{Gegenwart2008,PSSB:PSSB201300005,Wirth2016,NGCS}.  
Of particular interest are stimuli that control the (Kondo) coupling strength while leaving (nominal) valence states inert. 
In the case of heavy-fermion metals this allows, for instance, to directly explore Doniach's phase diagram\cite{DONIACH1977231}.
Isoelectronic tuning can be achieved, e.g., through external or chemical pressure.
In the latter case, elemental constituents are (at least partially) replaced with atoms having the same valence configuration but a different principle quantum number $n$.

In {\it binary} compounds, such isoelectronic substitutions directly control the unit-cell volume, as
$n$ typically correlates with the element's bonding radius.
Furthermore, calculations of the hybridization of the $f$-states with their surrounding display---in binary compounds---an
empirical one-to-one correspondence to the experimental Kondo temperature\cite{PhysRevMaterials.1.033802}.
Hence the notion that, both, external and chemical pressure manipulate the Kondo coupling $J_K$ primarily through the change in lattice constants.

\begin{figure}[!t]
{\includegraphics[clip=true,trim= 0 50 0 20, angle=0,width=.99\textwidth]{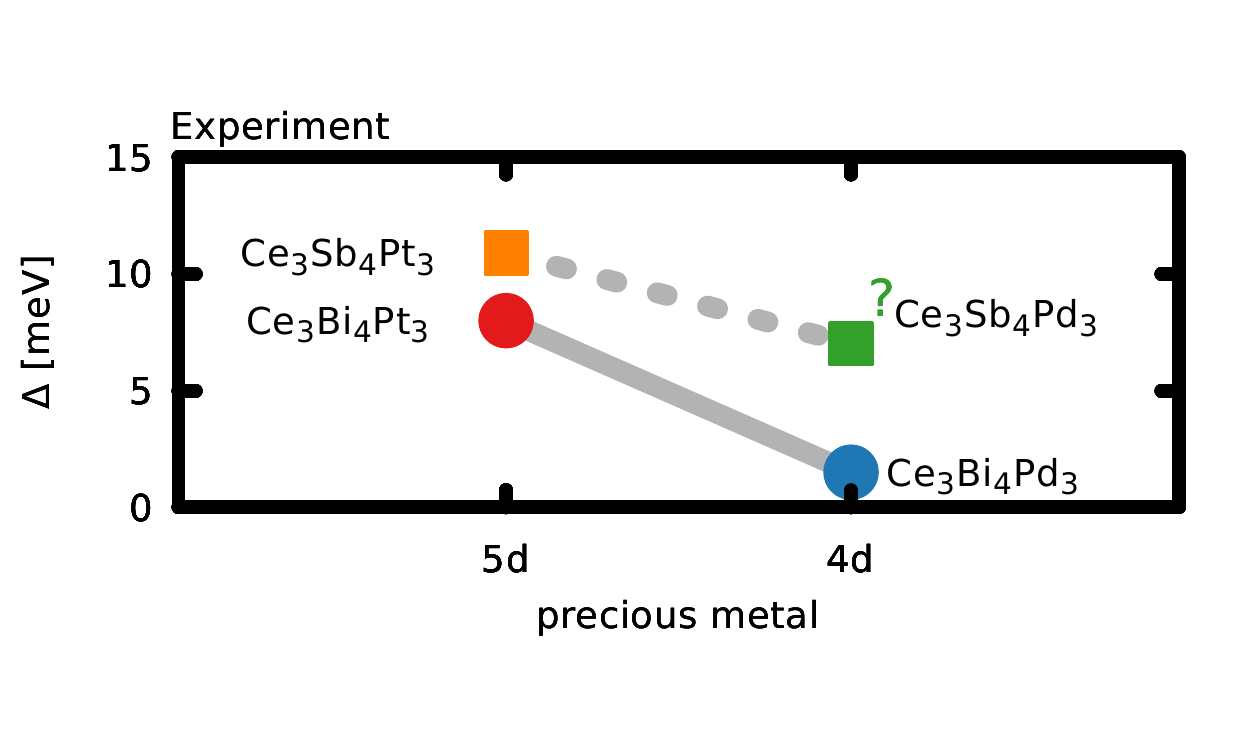}} 

\vspace{-0.5cm}
{\includegraphics[clip=true,trim= 0 25 0 0, angle=0,width=.99\textwidth]{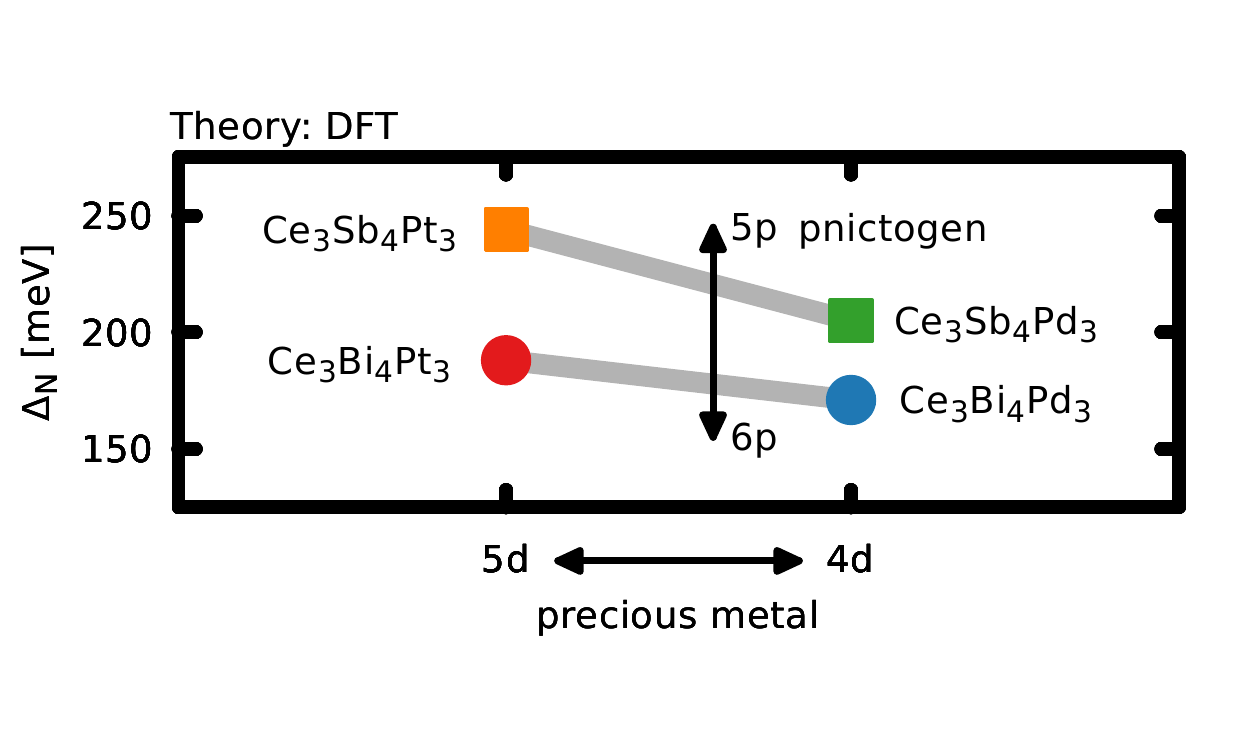}} 
\caption{{\bf Trends in the charge gap.}  
Top: experimental  gaps of our focus compounds (see also \fref{rhochi}); the value for Ce$_3$Sb$_4$Pd$_3$ is unknown.
Bottom: theoretical gaps within band-theory. 
Arrows indicate trends for pnictogen (Bi$\leftrightarrow$Sb) and precious metal (Pt$\leftrightarrow$Pd) substitution.
We seek to identify the microscopic control parameters that these substitutions engage.
Replacing the precious metal was shown\cite{PhysRevB.42.6842,PhysRevB.46.8067,Hermes2008} 
(is predicted, see \sref{sec:vol}) to be isovolume for the Bi (Sb) compounds.
Note that the gap $\Delta_N$ is measured from the Fermi level to the N-point, for reasons explained in the text.
}
\label{scheme}
\end{figure}

For more complex Kondo systems---ternary compounds and above---we here demonstrate that the unit-cell
volume (perhaps unsurprisingly) is {\it not} a good proxy for the size of the Kondo coupling.
Consider that in isoelectronic substitutions both
the shape and the radial extent of the wavefunction is different as the principle quantum number changes.
The will affect the hybridization and bonding not just with the atoms hosting the $4f$-orbitals but also between all other atoms.
Therefore it is not {\it a priori} clear whether isovalent substitutions modify (i) cohesive properties 
(largely dominated by hybridizations not involving the $f$-orbitals), (ii) the Kondo coupling $J_K$ 
(controlled by the hybridization of the $f$-electrons with their surrounding), or (iii) both of them.

Here, we explore how isoelectronic tuning affects the itineracy of $f$-electrons in the family of Kondo insulators\cite{FISK1995798,Riseborough2000,NGCS} 
Ce$_3$A$_4$M$_3$ (A=Bi,Sb, M=Pt,Pd), which is of high current interest\cite{PhysRevLett.118.246601,Dzsaber2018arxiv,jmt_CBP_arxiv,Dzsaber2019arxiv,Cao2019,Kushwaha2019,Campbell2019arxiv}. We consider both possible isovalent substitutions: that of the pnictogen and that of the precious metal.
We find that the difference between the Bi- and the Sb-derived materials 
are mainly driven by the unit-cell volume---a conventional case of isoelectronic tuning.
Interchanging the precious metal element in the Bi-compounds, on the other hand, does not change the unit-cell volume\cite{PhysRevB.42.6842,PhysRevB.46.8067,Hermes2008,PhysRevLett.118.246601}.
Nonetheless, we recently established\cite{NGCS} that the Kondo couplings of \cbp\ and \cbpd\ differ significantly and surmised that the main driver is the different {\it radial extent} of the 5d (Pt) and 4d (Pd) orbitals.
In the current work, we confirm this proposal and investigate the detailed interplay of volume, orbital extent, hybridization, and charge gaps within band-theory.
In other words, we elucidate the microscopic parameters that tune Ce$_3$A$_4$M$_3$ (A=Bi,Sb, M=Pt,Pd) along the arrows in \fref{scheme}(bottom).
Finally, by combining insights from our band-structure calculations with experimental magnetic susceptibilities, we predict 
the magnitude of mass renormalizations (\tref{tab1}):
We find (i) excellent agreement with recent many-body simulations for \cbp\cite{jmt_CBP_arxiv}; (ii)
the Pd-based system have significantly larger renormalizations than the Pt-ones; (iii)
the Sb-members have larger Kondo scales than their Bi-counterparts.
The latter highly motivates synthesizing and characterizing the so far unknown compound \cspd\ as well as the series
Ce$_3$Sb$_4$(Pt$_{1-x}$Pd$_x$)$_3$, in which the isovalent substitution can be studied at more accessible temperatures than for the Bi-series.

\section{Survey of experimental results and previous interpretations.}

\begin{figure}[!t]
{\includegraphics[clip=true,trim= 0 10 0 0, angle=0,width=0.99\textwidth]{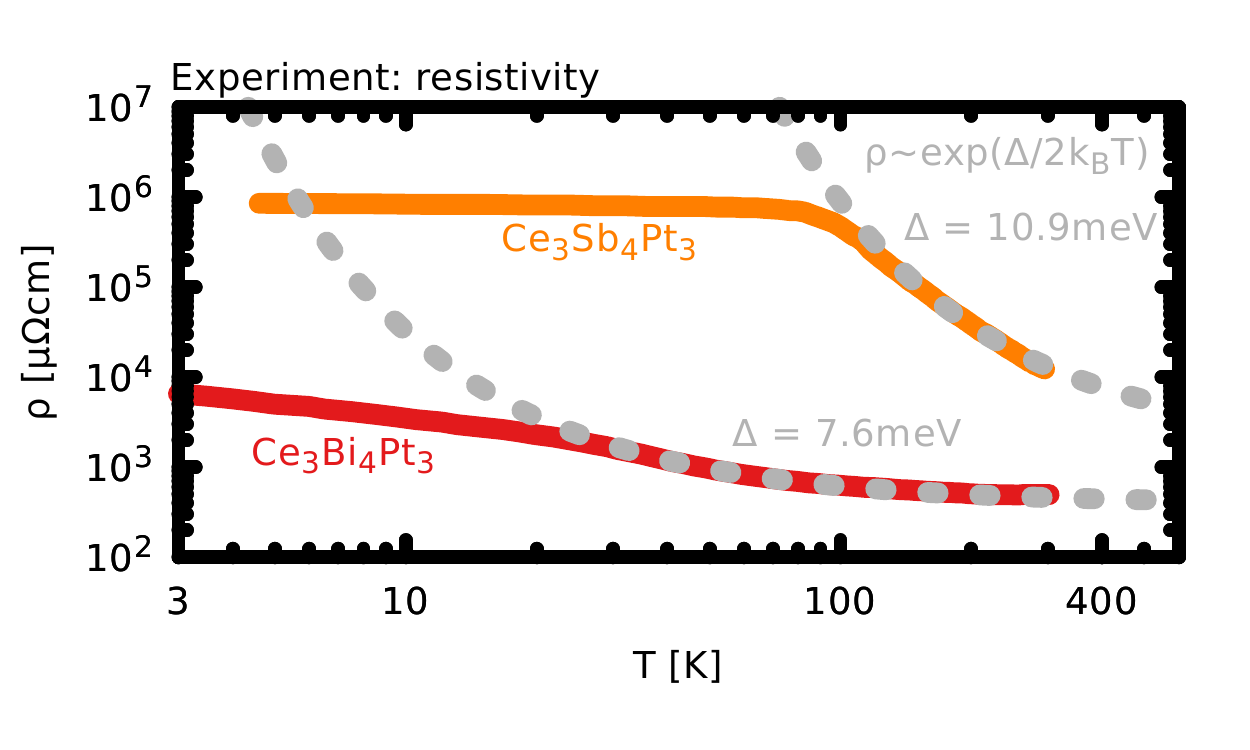}} 

\vspace{-0.25cm} 
{\includegraphics[clip=true,trim= 0 15 0 0, angle=0,width=.98\textwidth]{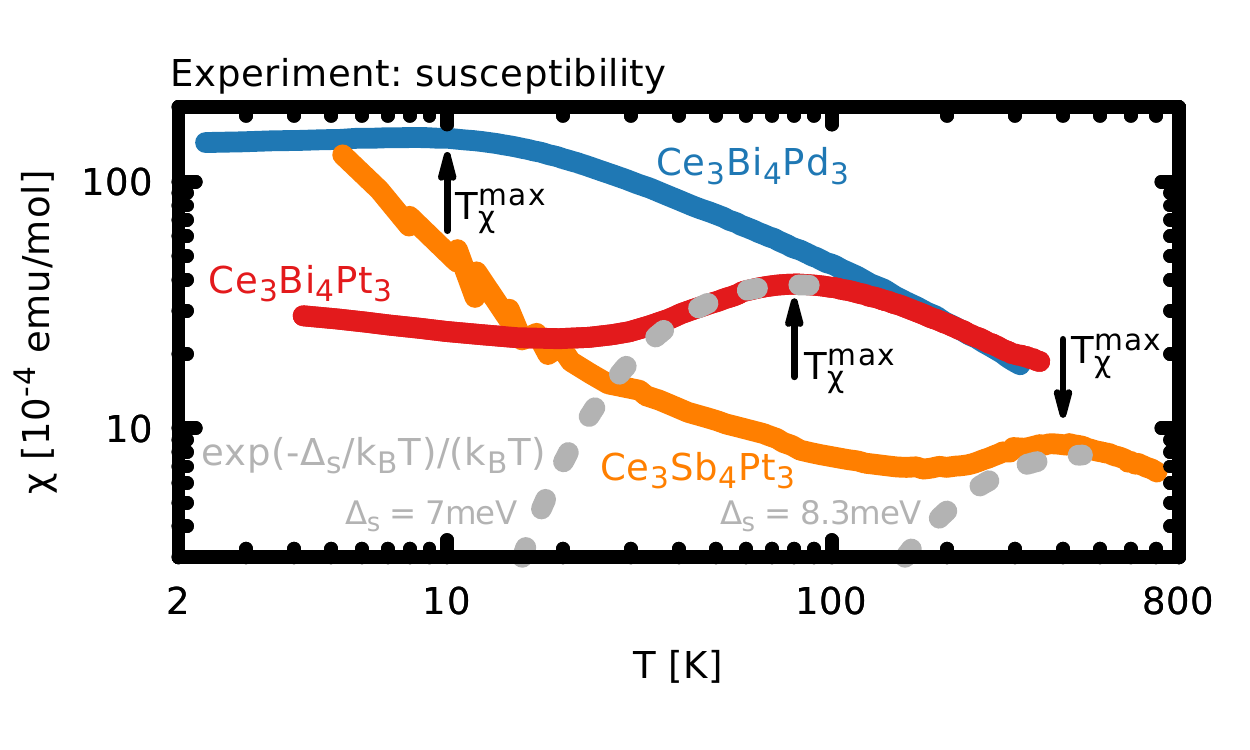}}  
\caption{{\bf Survey of experimental data.}  
Shown are the resistivities from Ref.~\onlinecite{Katoh199822} and Ref.~\onlinecite{PhysRevB.58.16057} (top panel) and magnetic susceptibilities
from Ref.~\onlinecite{PhysRevB.42.6842} and Ref.~\onlinecite{KASAYA1994534}
(bottom panel) of 
\cbp\ and \csp, respectively. The susceptibility of \cbpd\ is from Ref.~\onlinecite{PhysRevLett.118.246601}. Using phenomenological fits as indicated (in grey,
see also Ref.~\onlinecite{NGCS}), the charge and spin gaps are extracted. 
For $\chi$, the fitting ignores low-temperature Curie-Weiss contributions likely to derive from impurities.
The gap in \cbpd\ (no data shown) is estimated as $\sim 2$meV\cite{Kushwaha2019}.
}
\label{rhochi}
\end{figure}

We summarize pertinent experimental findings and their interpretation for the materials under consideration
(for a broader perspective see Refs.~\onlinecite{Riseborough2000,NGCS}):
As seen in the collection of resistivities (top) and magnetic susceptibilities (bottom) in \fref{rhochi},
the charge and spin gap, $\Delta$ and $\Delta_s$, and the crossover temperature, $T^{\hbox{\tiny max}}_\chi$, 
indicative of the onset of Kondo screening\cite{}, vary significantly in these materials: 
Interchanging of pnictogen and/or precious metal elements markedly tunes the energy scales.
The tendencies of the charge gap $\Delta$---on which we will focus our attention---are condensed in \fref{scheme}(top):
The Pt-based compounds are prime examples of Kondo insulators\cite{PhysRevB.42.6842,FISK1995798,KASAYA1991797,Riseborough2000}, with the Sb-member having a slightly larger charge gap than its Bi-counterpart. 
Exchanging Pt with Pd in \cbp\ diminishes the charge gap\cite{PhysRevB.42.6842,Katoh199822}, while the volume remains constant\cite{PhysRevB.42.6842,PhysRevB.46.8067,Hermes2008}.
The isovolume nature of the precious-metal substitution was recently confirmed by studying the continuous
series Ce$_3$Bi$_4$(Pt$_{1-x}$Pd$_x$)$_3$\cite{PhysRevLett.118.246601}.
The end member \cbpd\cite{Hermes2008} has been characterized as a semi-metal\cite{PhysRevLett.118.246601}, while more recent experiments\cite{Kushwaha2019}
extract a gap $\Delta\sim 2$meV from the Hall resistivity and the critical magnetic field.
Under the assumption of a direct correspondence between unit-cell volume and the Kondo coupling $J_K$, it was initially asserted\cite{PhysRevLett.118.246601} that $J_K$
is independent of $x$ in isovolume Ce$_3$Bi$_4$(Pt$_{1-x}$Pd$_x$)$_3$. It was then concluded\cite{PhysRevLett.118.246601} that 
the decrease of energy scales with increasing $x$ is driven by the change in the spin-orbit coupling, owing to the mass differences of Pt and Pd---an exciting perspective in view of the recent proposals of topological Kondo insulators\cite{Dzero2016,Chang2017}.
Here, instead, we argue in favour of a substantial evolution of the Kondo coupling in the series Ce$_3$Bi$_4$(Pt$_{1-x}$Pd$_x$)$_3$.
The compound \cspd\ has---to the best of our knowledge---not yet been synthesized.

\begin{figure}[!t]
{\includegraphics[clip=true,trim= 0 30 0 0, angle=0,width=.98\textwidth]{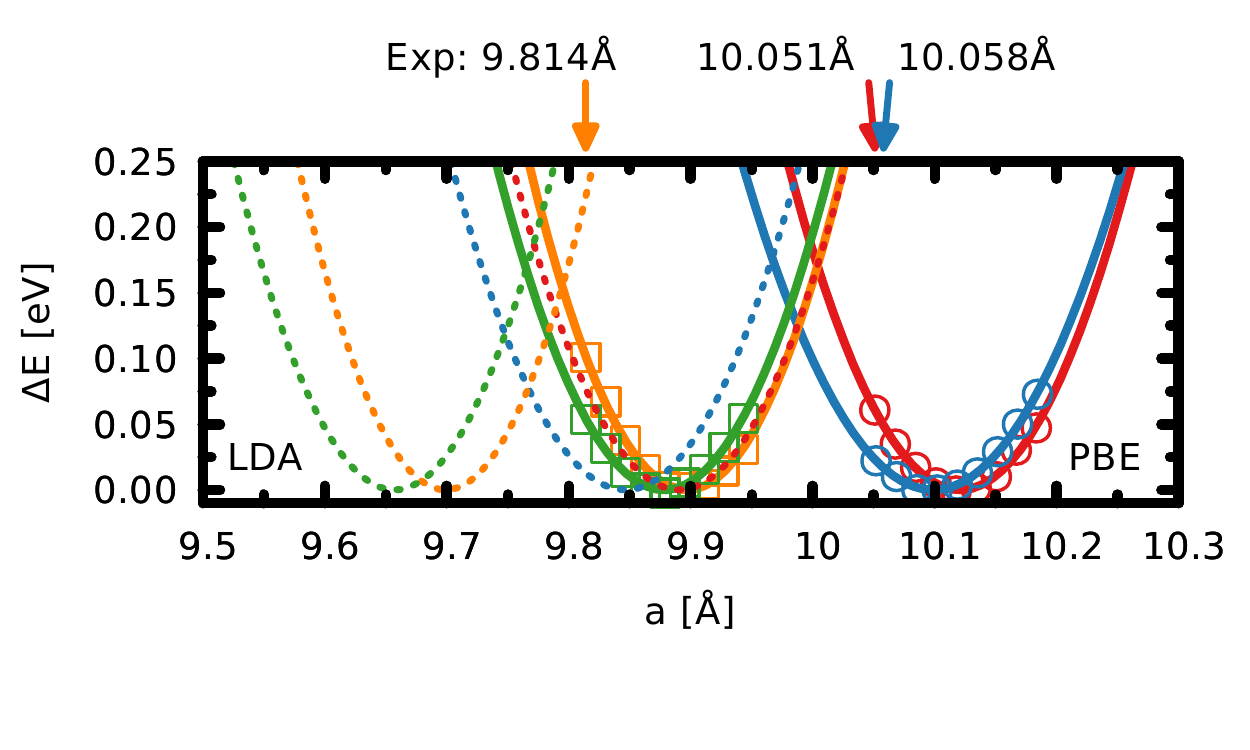}} 
\caption{{\bf Experimental and theoretical lattice constants}. Shown is the relative energy difference $\Delta E$ for varying lattice constant $a$,
using the PBE (symbols and solid lines) and the LDA (dashed lines) potential, where lines are parabolic fits. Arrows indicate experimental lattice constants
(see text). The colours are chosen as follows:
 \cbp\ \red{\ding{108}}, \cbpd\ \blue{\ding{108}}, \csp\ \yellow{\ding{110}}, \cspd\ \green{\ding{110}}.
In that order, the computed equilibrium lattice constants are 10.12, 10.10, 9.90, 9.88\AA\ with PBE and
9.89, 9.85, 9.70, 9.66\AA\ using LDA.
}
\label{volume}
\end{figure}


\begin{figure*}[!t]
\begin{center}
  \sidesubfloat[]{\includegraphics[clip=true,trim= 5 25 0 25, angle=0,width=0.45\textwidth]{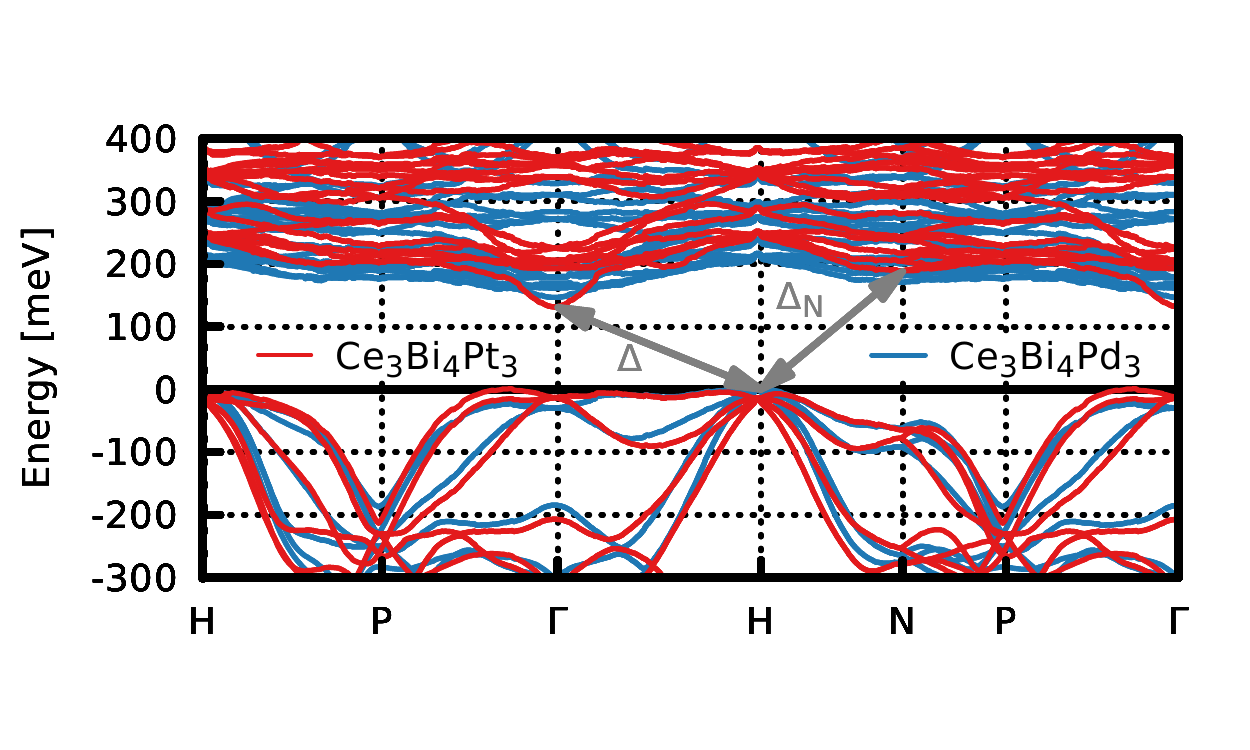}\label{bands:sub1}}\quad%
  \sidesubfloat[]{\includegraphics[clip=true,trim= 5 25 0 25, angle=0,width=0.45\textwidth]{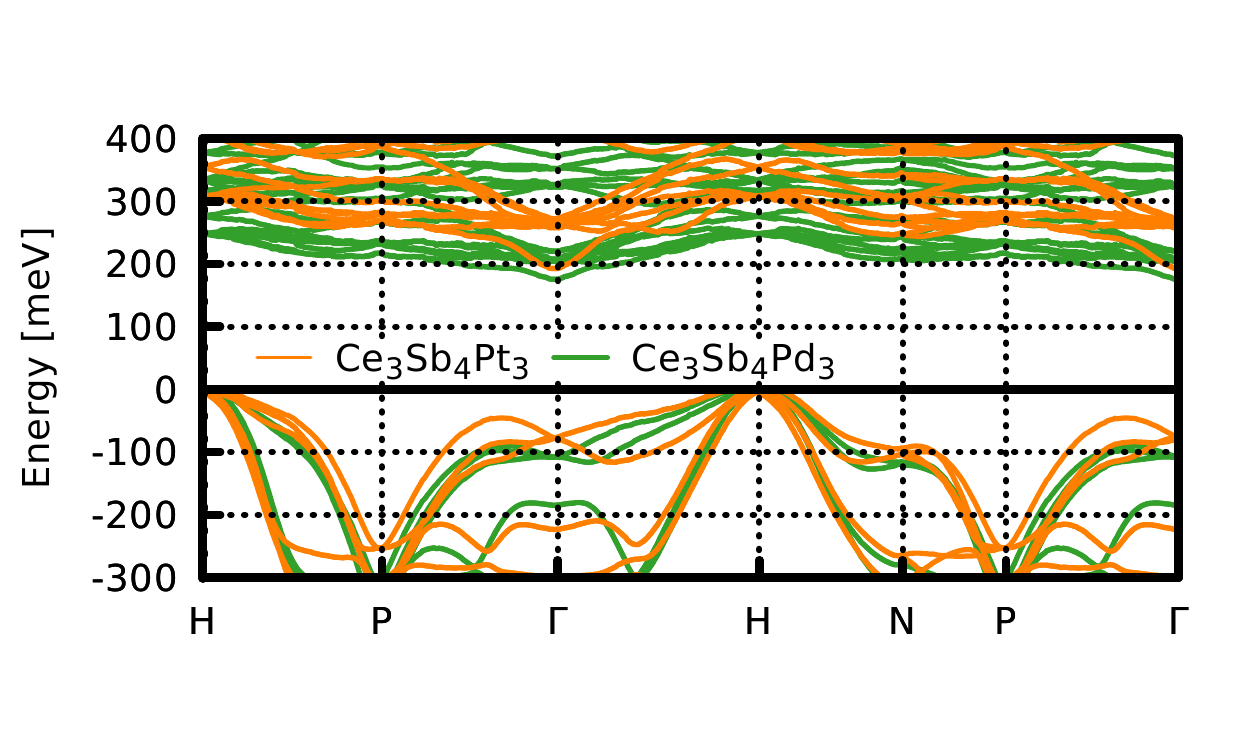}\label{bands:sub2}}%
  \caption{{\bf DFT band-structures.} \cbp\ \& \cbpd\ (a) and \csp\ \& \cspd\ (b). 
On the left are indicated the fundamental indirect gap $\Delta$ and the gap $\Delta_N$ used as a proxy for
the apparent gap in the density of states in \fref{dos}(a).}\label{bands}
\end{center}
\end{figure*}

\section{Crystal-structure and Methods.}

The compounds Ce$_3$A$_4$M$_3$ (A=Bi,Sb, M=Pt,Pd) are a family of cubic intermetallics that 
crystallize in the body-centred Y$_3$Sb$_4$Au$_3$ structure\cite{Dwight:a14686} which has the non-symmorphic space-group I-43d, two formula units per unit-cell and a
lattice constant  $a=10.051$-$10.058$\AA \cite{PhysRevB.42.6842,PhysRevB.46.8067,Hermes2008,PhysRevLett.118.246601} ($a=9.814$\AA \cite{PhysRevB.60.5282}) for \cbp\ and \cbpd\ (\csp).
In our band-structure calculations we used $a=10.051$\AA\ ($a=9.814$\AA) for  Ce$_3$Bi$_4$M$_3$ (Ce$_3$Sb$_4$M$_3$), as motivated below in \sref{sec:vol}.
In all four compounds, the Ce atoms occupy in the Wyckoff position 12a (3/8, 0, 1/4), Pt/Pd sits at site 12b (7/8, 0, 1/4), and Bi/Sb is found at (u,u,u). 
Here, we have followed Ref.~\onlinecite{doi:10.1143/JPSJ.62.2103} and used the ideal position $u=1/12$, 
which is very close to $u=0.0857$ found for \cbp\cite{PhysRevB.46.8067} and $u=0.084$ of \cbpd\cite{Hermes2008}.

In this setting, we performed electronic structure calculations using the wien2k package\cite{wien2k}. We employed the PBE functional,
and included spin-orbit coupling (SOC) in the scalar relativistic approximation for all atoms. 
Volume optimizations were performed with $12^3$ reducible k-points, self-consistency for experimental lattice constants 
was reached with a k-mesh of $20^3$ points.
We use the auxiliary Kohn-Sham spectrum as proxy for the electronic structure of the studied materials.
Statements about inter-atomic hybridizations are based on: (a) maximally localized Wannier functions\cite{RevModPhys.84.1419} constructed 
in the total angular momentum $J$-basis for the subspace of Ce-$4f$, A-$p$ and M-$d$ orbitals with Wannier90\cite{wannier90} via wien2wannier\cite{wien2wannier} and
using a tool of J.\ Fernandez Afonso\cite{JuanPhD} to rotate the local spin coordinate system;
(b) a local projection formalism due to Haule \etal \cite{PhysRevB.81.195107}.
The investigation of the electronic structure of the Pt-members has been pioneered by Takegahara \etal\ \cite{doi:10.1143/JPSJ.62.2103}. The
band-structures and density of states of the two Bi-compounds have been previously shown in the review article Ref.~\onlinecite{NGCS}.

\subsection{Comment: Band-theory for Kondo insulators}

Given that effective one-particle theories fail in most respects for systems with $f$-electrons, a few comments are in order.
Indeed, the presence of quasi-localized states makes the use of genuine many-body methods a prerequisite for an adequate description,
see, e.g., Refs.~\onlinecite{PhysRevLett.87.276404,amadon:066402,PhysRevB.81.195107,jmt_cesf,PhysRevLett.112.106407,Shick2015,Wissgott2016,PhysRevB.98.075129,jmt_CBP_arxiv,Cao2019,PhysRevB.100.035121}.
Nonetheless, conventional electronic structure theory can give meaningful insights into at least the trends of properties 
in families of $f$-electron compounds\cite{0034-4885-79-12-124501,PhysRevMaterials.1.033802,Hafiz2018}.

This is particularly true for (Ce-based) Kondo insulators:
In the language of the Kondo lattice model (KLM), the charge gap in Kondo insulators is interaction-driven through the Kondo coupling $J_K$.
However, this spin-fermion interaction is not a fundamental force, but a quantity that emerges in the renormalization group-like flow to a low-energy description.
Indeed, it is well known that the KLM can be obtained
as a limiting case of the periodic Anderson model (PAM), which contains the Hubbard $U$ interaction as a remnant of the fundamental Coulomb force:
The strong $U$-coupling limit of the PAM yields the weak $J_K$-coupling regime of the KLM with
 $J\propto V^2/U$%
\footnote{Actually, there exists an exact mapping of the PAM to the KLM beyond the Kondo regime\cite{PhysRevB.65.212303}.}.
The above illustrates that (bare) hybridizations $V$ between conduction electrons and the $f$-level and the Hubbard interaction $U$, together, generate the Kondo coupling. 
Note that, while the KLM is metallic for vanishing interaction, $J=0$, in the absence of Hubbard $U$-interactions, the (symmetric, half-filled) PAM is a band-insulator.
In this sense, the Kondo insulating gap in the PAM is not interaction-driven.

Similar to the non-interacting PAM, band-theory finds Ce$_3$A$_4$M$_3$ to be covalent band-insulators.
Band-theory of course grossly overestimates the hybridization $V$ of $f$-states with their surroundings.
Yet, it is sensitive to changes in the Lanthanide environment (external pressure, isoelectronic substitutions)
and can therefore qualitatively capture modifications in the essential inter-atomic hybridizations and the Kondo coupling.
Hence,  {\it trends} in the hybridization gap of Ce$_3$A$_4$M$_3$ within DFT
can  be used as a gauge for the Kondo coupling.

Let us note that this situation is facilitated by (i) the  nominal $4f^1$ configuration in Ce-based compounds\cite{PhysRevB.27.7330,PhysRevB.86.081105,STRIGARI201556}, as multiplet effects are less crucial 
than in systems with more electrons in the $f$-shell%
\footnote{See, however, e.g., Refs.~\onlinecite{doi:10.1080/01411590701228703,PhysRevLett.112.106407} for multiplet effects in Ce-compounds under external pressure.}%
; (ii) the $f$-electrons in Ce-materials are more delocalized than, e.g.,
in the hole-analogues based on trivalent Yb; (iii) contrary to Mott insulators, the self-energy of Kondo insulators is benign near the Fermi level\cite{NGCS},
yielding a renormalized band-structure below the Kondo temperature.

\section{Results \& Discussion.}

\subsection{Volume optimization.}\label{sec:vol}

We assess the isovolume nature of Ce$_3$Bi$_4$(Pt$_{1-x}$Pd$_x$)$_3$ by optimizing the volume of the end members
($x=0,1$) within DFT for fixed fractional atomic positions.
As can be seen in \fref{volume} the theoretical lattice constants of \cbp\ and \cbpd---identified as the minima in the energy curves---are indeed very similar.
We find, using PBE and the Pt-compound as reference: $\Delta a/a=0.2 \%$. Experiments find instead that \cbpd\ is minutely larger than \cbp: $\Delta a/a= -0.07 \%$\cite{PhysRevLett.118.246601}.
We repeat the same calculations for the pair Ce$_3$Sb$_4$M$_3$ (M=Pt,Pd). We find that exchanging Bi with Sb yields significantly smaller compounds.
Crucially, the exchange Pt$\leftrightarrow$Pd leads again only to very small changes: $\Delta a/a= 0.2 \%$.%
\footnote{All relative changes are slightly larger when using LDA instead of PBE.}
We interpret the overall small changes in lattice constant, incurred upon precious-metal substitution, to warrant the label 'isovolume'.
For Ce$_3$Bi$_4$M$_3$ this supports previous experimental findings\cite{PhysRevB.42.6842,PhysRevB.46.8067,Hermes2008,PhysRevLett.118.246601}, while the isovolume nature of
Ce$_3$Sb$_4$(Pt$_{1-x}$Pd$_x$)$_3$ is a prediction.

Let us remark that replacing Pt with Pd is not affecting the lattice constant is not a generic behaviour.
Indeed, the difference in lattice constant between CePtSn\cite{TAKABATAKE1993108}
and CePdSn\cite{PhysRevB.40.2414}, amounting to $\Delta a/a=0.8\%$,
is experimentally an order of magnitude larger than in Ce$_3$Bi$_4$(Pt$_{1-x}$Pd$_x$)$_3$.
We can speculate that in the more complex unit-cell of Ce$_3$A$_4$M$_3$ (in which, e.g., every ``M'' has 4 nearest ``A'' neighbours)
small atomic differences 
can be better compensated than in
CeMB
(where each ``M'' has only 3 neighbouring ``B''). 

We further note that---in line with common experience---PBE(LDA) over(under)estimates the lattice constant.
In the following electronic structure calculations we have therefore settled to employing experimental values:
we use the lattice constant of \cbp\ for Ce$_3$Bi$_4$M$_3$, and the one of \csp\
for Ce$_3$Sb$_4$M$_3$, where M=Pt,Pd.

\subsection{Band-structures.}

\begin{figure*}[!t]
  \begin{center}
\sidesubfloat[]{\scalebox{1.}{\includegraphics[clip=true,trim= 0 25 0 20,angle=0,width=.45\textwidth]{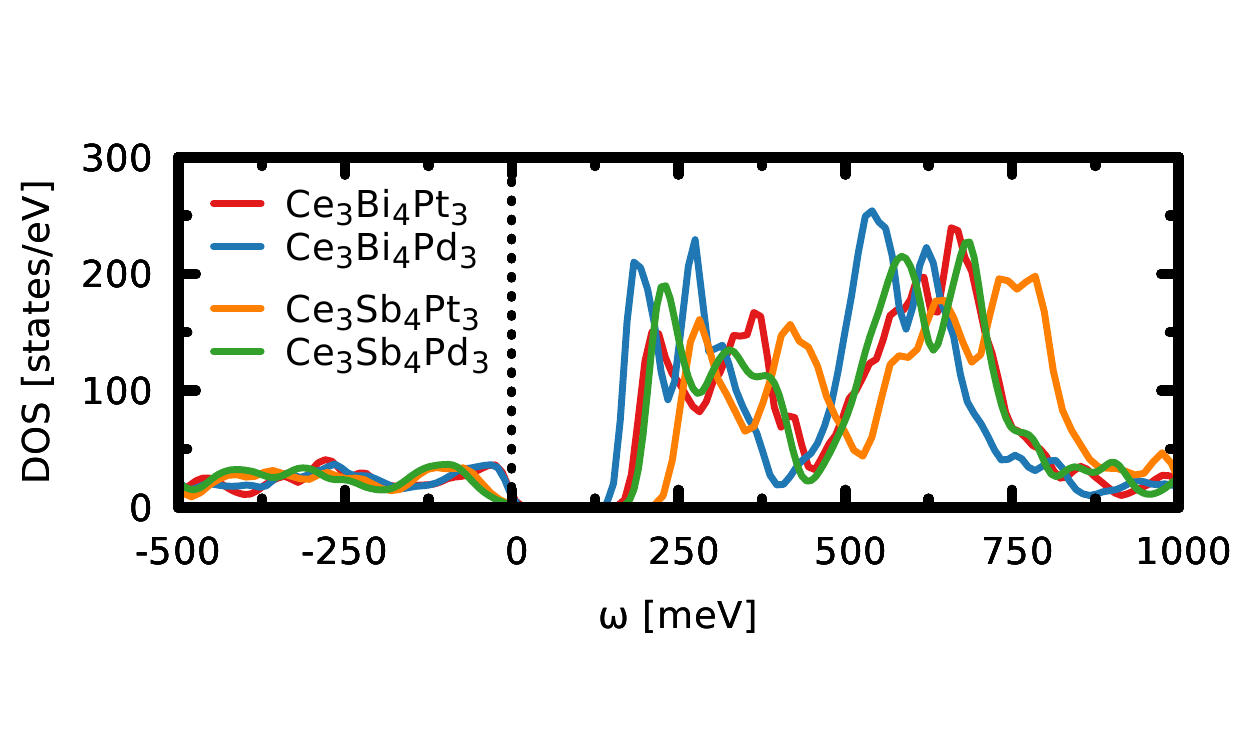}}\label{dos:sub1}}\quad%
\sidesubfloat[]{\scalebox{1.}{\includegraphics[clip=true,trim= 0 25 0 20,angle=0,width=.45\textwidth]{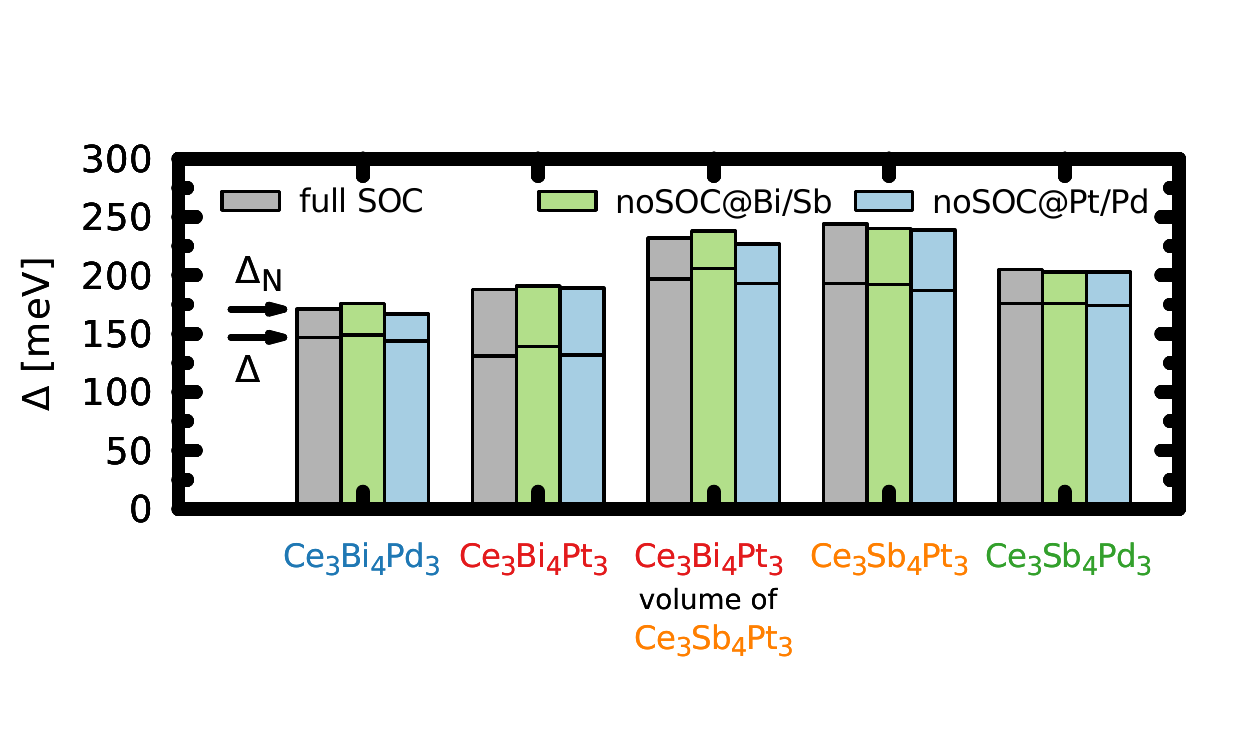}}\label{dos:sub2}} 
      \caption{{\bf Density of states and charge gaps.} 
			Panel (a) displays the density of states (DOS) for all four materials. Panel (b) collects the fundamental indirect gap $\Delta$ occurring between the valence band maximum near the H-point and the conduction band minimum at $\Gamma$. Also shown are the larger values $\Delta_N$ that
			are measured between the valence maximum and onset of conduction states at the N-point. Owing to the small weight of conduction states at $\Gamma$, $\Delta_N$ is more representative for the trends of the gap as apparent in the DOS, see also \fref{bands}.
			As illustrated by selectively turning off the  spin-orbit coupling (SOC)  for the pnictogen and precious metal atoms, it is seen that within DFT the SOC does not control
			the overall trends within this family of compounds. Additionally shown are gap values of fictitious \cbp\ that uses the lattice constant/volume of \csp.
}
      \label{dos}
      \end{center}
\end{figure*}

\fref{bands} displays the band-structures of all Ce$_3$A$_4$M$_3$ compounds considered here, while \fref{dos}(a) shows the corresponding
density of states (DOS). 
For a given pnictogen ``A'', the valence states are rather similar, irrespective of the precious metal ``M''.
Changing the pnictogen, results in perceptible changes of the valence states, in particular near the Brillouin zone centre.
Overall, the valence states are---congruent with photoemission spectroscopy\cite{0295-5075-41-5-565,Takeda1999721}---comparatively dispersive
as they are dominantly of $pd$-character. 
Indeed, as expected with only one nominal $4f$-electron, quasi-atomic narrow bands are prevalent only above the Fermi level,
where also the DOS exhibits pronounced peaks.

In this work, we are mostly interested in the trends of the charge gap under isovalent substitutions.
The gaps $\Delta$ extracted from the band-structures are reported in \fref{dos}(b).
All gaps are indirect and spread, for most compounds, from around the H-point to the zone centre $\Gamma$.
An exception is \cbp, where the top of the valence bands emerges between the P and the $\Gamma$-point.
Another speciality of \cbp\ is the fact that the $\Gamma$-point conduction-band minimum is particularly pronounced.
While the DOS---dominated by the $f$-spectral weight---suggests that the charge gap {\it increases} when substituting Pd for Pt, the fundamental gap of \cbpd---controlled
by the dispersive band at $\Gamma$---is actually
larger than in \cbp\ within DFT. Since we believe the physics of these compounds to be dominated by the $f$-electrons, we introduce---as a proxy for the
trends in the $f$-band position---the quantity $\Delta_N$ which measures the gap between the Fermi level and onset of the conduction states at the $N$-point.
To disentangle putative control parameters of the charge gap, we report in \fref{dos}(b) also the values of $\Delta$ and $\Delta_N$ when  turning off the SOC
on selected atoms. Furthermore, so as to demonstrate the effect of the unit-cell volume onto the charge gap, we also show results for a fictitious variant of \cbp\ that has the lattice constant of \csp.

\subsection{Analysis of electronic structure}

In this discussion section, we asses the influence of the spin-orbit coupling, the unit-cell volume, and the
radial extent of orbitals onto the charge gap.

\subsubsection{Spin-orbit coupling}

As seen in \fref{dos}(b), the SOC is---within DFT---irrelevant for the qualitative trends of the charge gap in this family of compounds:
Turning selectively off the SOC for Pt in \cbp---to simulate the change in atomic mass of the precious metal atoms---does {\it not} result in a band gap comparable to isovolume \cbpd.
In fact, this procedure yields a $\Delta$ and $\Delta_N$ that surpass the values obtained with SOC on all atoms.
Hence, the difference in masses of Pt and Pd
and the resulting change in the magnitude of relativistic effects is likely not 
the driver of the Kondo insulator to semi-metal transition in 
Ce$_3$Bi$_4$(Pt$_{1-x}$Pd$_x$)$_3$ of Ref.~\onlinecite{PhysRevLett.118.246601} (see also the discussion in Ref.~\onlinecite{NGCS}).
Also, eliminating the SOC of Bi in \cbp---to simulate a smaller pnictogen mass when going to \csp---cannot bridge 
the disparity of band-gaps in these two materials. 
The spin-orbit coupling propels the properties neither along the vertical (Bi$\leftrightarrow$Sb) nor the horizontal (Pt$\leftrightarrow$Pd) arrow in \fref{scheme}.

\subsubsection{Hybridization nature of the gap}

A far more common tuning parameter in heavy fermion systems is the unit-cell volume.
While the precious metal substitution series Ce$_3$Bi$_4$(Pt$_{1-x}$Pd$_x$)$_3$ is isovolume\cite{PhysRevB.42.6842,PhysRevB.46.8067,Hermes2008,PhysRevLett.118.246601},
changing the pnictogen atom from Bi to Sb decreases the lattice constant by 2\%\cite{PhysRevB.42.6842,PhysRevB.46.8067,PhysRevB.60.5282}.
To isolate the effect of the lattice, we report, in \fref{dos}(b), the charge gap of fictitious \cbp\ where the lattice constant has been artificially shrunken to the value realized in \csp. 
With respect to true \cbp, we find a significantly increased gap. In fact, the change in the unit-cell volume accounts for almost 80\%
of the gap difference between \csp\ and \cbp.
This finding identifies the unit-cell volume as the leading protagonist for the trends between the Bi and the Sb {\it pairs} of materials.

That the lattice has such a large effect onto the electronic structure finds its origin in the hybridization nature of the gap in these systems.
Indeed, as has been demonstrated in Ref.~\onlinecite{NGCS} it is the hybridization of the Ce-$4f$ orbitals to their surrounding that accounts for the gap opening:
Treating the Ce-$f$ (J=5/2) as core-states results in metallic ground-states, in analogy to a periodic Anderson model with vanishing hybridization between localized $f$ and conduction states.

\begin{figure*}[!t]
  \begin{center}
\sidesubfloat[]{\scalebox{1.}{\includegraphics[clip=true,trim= 0 25 0 25,angle=0,width=.45\textwidth]{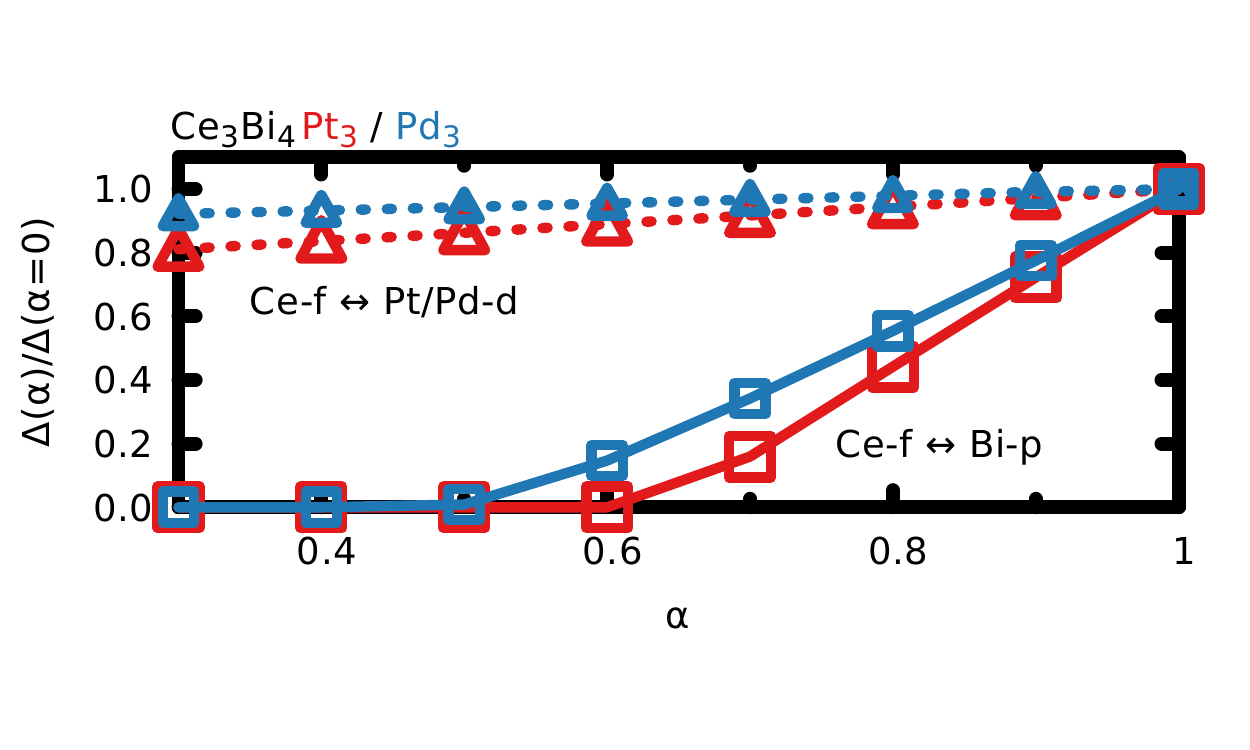}\label{scalehyb:sub1}}}\quad%
\sidesubfloat[]{\scalebox{1.}{\includegraphics[clip=true,trim= 0 25 0 25,angle=0,width=.45\textwidth]{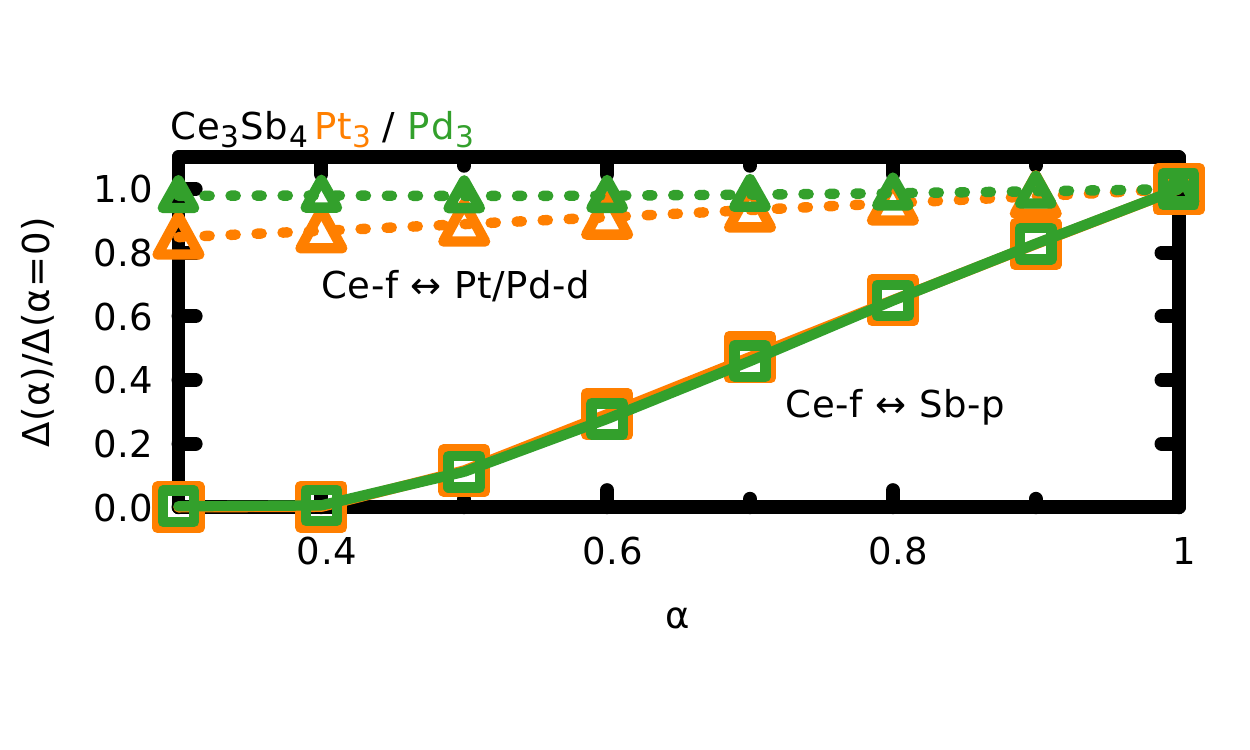}\label{scalehyb:sub2}}}
      \caption{{\bf Hybridization nature of the charge gap.}  
			Shown is the relative magnitude of the gap $\Delta$
			when scaling the hybridization between the Ce-$4f$ and the M-$d$ (M=Pt, Pd), or A-$p$ (A=Bi,Sb) orbitals in a maximally localized Wannier basis
			with a factor $\alpha\leq1$ for Ce$_3$Bi$_4$M$_3$ (a) and Ce$_3$Sb$_4$M$_3$ (b). 
			}
      \label{scalehyb}
      \end{center}
\end{figure*}

Here, we want to further distinguish the individual contributions of atoms/orbitals to the hybridization gap $\Delta$.
To this end we constructed maximally localized Wannier functions (see method section) and expressed the projected DFT Hamiltonian in that basis.
\fref{scalehyb} shows what happens to the relative magnitude of the gap in (a) Ce$_3$Bi$_4$M$_3$ and (b) Ce$_3$Sb$_4$M$_3$ when the hybridization (off-diagonal elements in the Hamiltonian) between the 
Ce-$4f$ orbitals and the M$-d$ (M=Pt,Pd) or the Bi/Sb$-p$ orbitals is scaled with a factor $\alpha\leq 1$.%
\footnote{See also Ref.~\onlinecite{NGCS} where such a construction helped distinguishing the gap nature in covalent FeSi and ionic LaCoO$_3$.}
We find the gap to be most sensitive to changes in the hybridization between the $4f$-orbitals of Ce and the $p$-orbitals of Bi and Sb.
For the Sb compounds, a smaller scaling factor $\alpha$ is needed to collapse the gap,
congruent with Ce$_3$Sb$_4$M$_3$ having larger hybridizations than their Bi-counterparts.
The gap is much less sensitive to hybridizations between the Ce-$f$ and 
the precious metal M-$d$-orbitals. We note that for both A=Bi and A=Sb the gap is more quickly diminished with $\alpha$ if the precious metal is M=Pt.
The relative suppression of $\Delta(\alpha)$ when scaling the precious metal-hybridization is similar irrespective of the type of the pnictogen ``A''.
This finding supports the assertion that the  anticipated diminishing of the gap in Ce$_3$Sb$_4$(Pt$_{1-x}$Pd$_x$)$_3$ with $x$ is
not volume driven---akin to what happens in Ce$_3$Bi$_4$(Pt$_{1-x}$Pd$_x$)$_3$.

\subsubsection{Radial extent of orbitals}

\begin{figure*}[!t]
\sidesubfloat[]{\scalebox{1.}{\includegraphics[clip=true,trim= 90 40 98 0,angle=0,width=.35\textwidth]{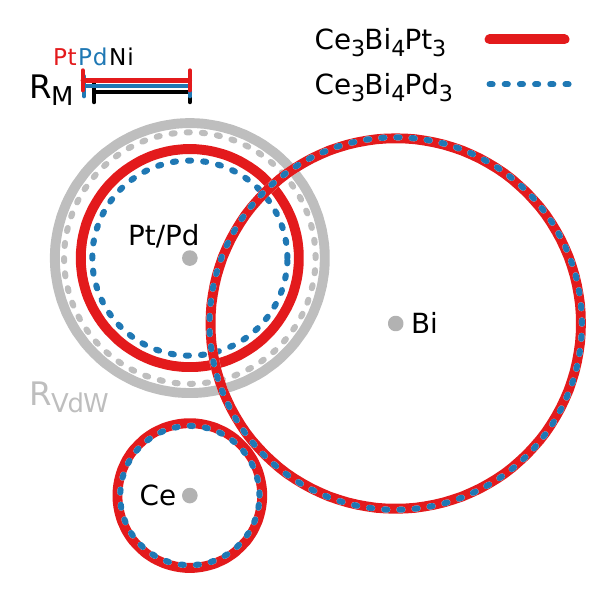}}}\qquad%
\sidesubfloat[]{\scalebox{1.}{\includegraphics[clip=true,trim= 90 40 98 0,angle=0,width=.35\textwidth]{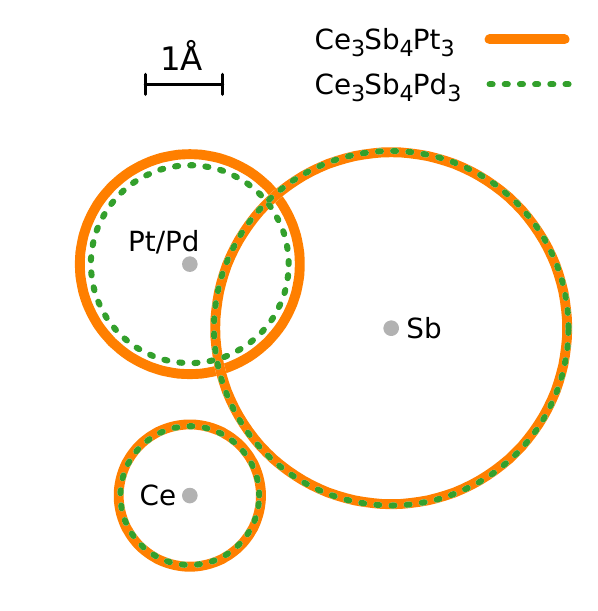}}} 

\hspace{0.75cm}

\sidesubfloat[]{\scalebox{1.}{\includegraphics[clip=true,trim= 85 0 95 0,angle=0,width=.2\textwidth]{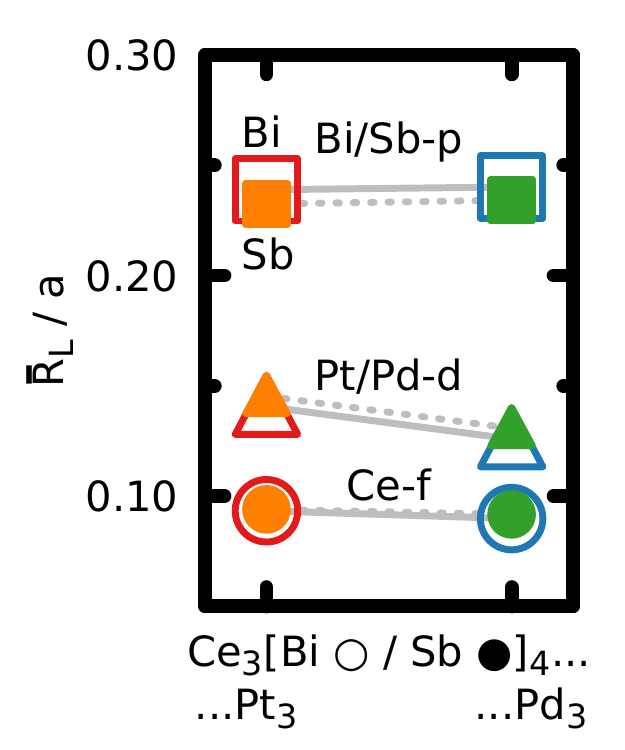}}}\quad%
\sidesubfloat[]{\scalebox{1.}{\includegraphics[clip=true,trim= 85 0 95 0,angle=0,width=.2\textwidth]{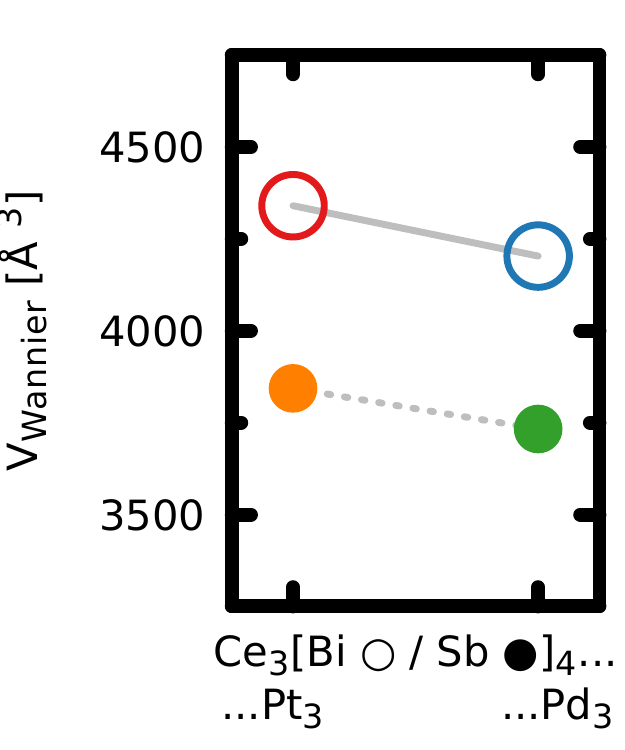}}}\quad%
\sidesubfloat[]{\scalebox{1.}{\includegraphics[clip=true,trim= 85 0 95 0,angle=0,width=.2\textwidth]{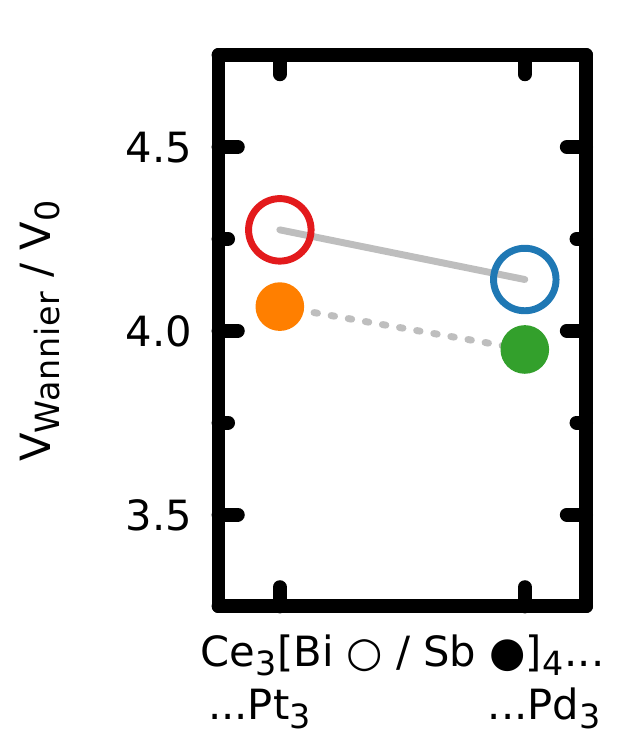}}}\quad%
\sidesubfloat[]{\scalebox{1.}{\includegraphics[clip=true,trim= 85 0 95 0,angle=0,width=.2\textwidth]{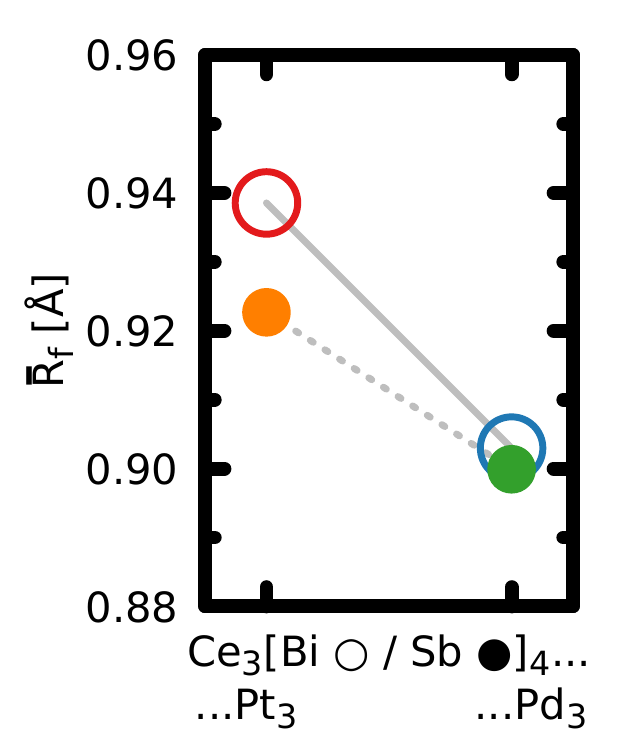}}}\quad%
      \caption{{\bf The radial extent of orbitals.} 
			Panels (a-b) visualize (to scale) the average radial extent (square-root of the Wannier spread) of the Ce-$f$, M-$d$ (M=Pt,Pd), A-$p$ (A=Bi,Sb)
			orbitals in relation to inter-atomic nearest-neighbour distances.
			Also shown in (a) is the Van der Waals radius $R_{VdW}$ of Pt (1.75\AA) and Pd (1.63\AA)\cite{Bondi1964} in solid (dashed) gray,
			as well as the metallic radii $R_M$ of Pt (1.39\AA), Pd (1.37\AA), and Ni (1.24\AA)\cite{Greenwood1997}.
			Panel (c) display the average radial extent $\overline{R}_L$  {\it relative} to the respective lattice constant $a$. 
			Panel (d) shows the summed volume $V_{Wannier}$ of all Wannier orbital-spheres of radius $R_L$ in the unit-cell.
			Panel (e) shows the same as (c) but divided by the unit-cell volume $V_0$.
			Panel (f) displays the average radial extent $\overline{R}_f$ of the Ce-$4f$ orbitals; the relative extent $\overline{R}_f/a$ is shown in \fref{scheme_new}.
			}      \label{Rfig}
\end{figure*}

Now we will disentangle changes in the inter-atomic hybridization mediated by (i) the unit-cell volume and (ii) the radial extent of pertinent orbitals.
To this end we and analyse the orbitally ($l$-)resolved spread $\Omega_l=\bra{l} R^2\ket{l}-\bra{l} R\ket{l}^2$ of the Wannier functions,
the sum of which is minimized in the maximally localization procedure\cite{RevModPhys.84.1419}.
We define the mean extent of a type $L$ of orbitals as the average of the square-root
of the $l$-resolved Wannier spread: $\overline{R}_L=1/N_L\sum_{l\in L}\sqrt{\Omega_l}$.
For example, in case of the Ce-$4f$ orbitals, i.e. $L=4f$, $l$ runs over the $N_L=14$ spin-orbitals per Ce-atom.
\fref{Rfig} visualizes the thus measured size of orbitals  to scale with nearest-neighbour atomic distances for 
(a) Ce$_3$Bi$_4$M$_3$,
(b) Ce$_3$Sb$_4$M$_3$
where M=Pt,Pd. Panel (c) displays the trends of $\overline{R}_L$ for precious metal and pnictogen substitutions
relative to the materials' lattice constant $a$.

First, we discuss the trends upon interchanging Pt $\leftrightarrow$ Pd.
Congruent with the precious metals' van der Waals\cite{Bondi1964} and metallic\cite{Greenwood1997} radii
indicated in \fref{Rfig}(a), we find the 
Pd-$4d$ orbitals to be smaller than the Pt-$5d$ ones, irrespective of the pnictogen. 
The Bi and Sb-$p$ orbitals instead are slightly larger when the precious metal is palladium, which in part compensates
for the chemical pressure of the smaller Pd.
It is thus suggestive to ascribe this expansion of the pnictogen to play a role 
in the Ce$_3$Bi$_4$(Pt$_{1-x}$Pd$_x$)$_3$ series being  isovolume.
Nonetheless, computing the volume of all ``Wannier spheres'' (of radii $\overline{R}_L$) in the unit-cell,
$V_{Wannier}=4/3 \pi \times \left(6\times 14{\overline{R}_f}^3+4\times 6{\overline{R}_p}^3+3\times 10{\overline{R}_p}^3\right)$,
we find for \cbp\ ($V_{Wannier}=4340\AA^3$) a value still larger by 3\% than for \cbpd\ ($V_{Wannier}=4203\AA^3$), see \fref{Rfig}(d).
The relative difference is similar for the Sb-based compounds
 \csp\ ($V_{Wannier}=3843\AA^3$) and \cspd\ ($V_{Wannier}=3732\AA^3$).
This suggests that the crystal-structure is held together dominantly by the (much larger) pnictogen $p$-orbitals, while the contribution to bonding
from the precious metal $d$-orbitals is sub-leading.
The small dependence of the hybridization gap on the Ce-$4f$ to Pt/Pd-$d$ hopping (see \fref{scalehyb}) supports this statement.

The above is further substantiated when looking at the exchange of the pnictogen: Bi $\leftrightarrow$ Sb.
There, we find the smaller orbitals of the Sb-based compounds to be in line with the
smaller unit-cell volume as compared to their Bi-counterparts (lattice constants: $10.051\AA$ for Ce$_3$Bi$_4$M$_3$ and $9.814\AA$ for \csp, see \sref{sec:vol}).
Indeed, the Wannier volume (as defined above) decreases by more than 10\% when replacing Bi with Sb, see \fref{Rfig}(d).
Note, that this trend with (pnictogen) {\it chemical} pressure is opposite to the generic behaviour of Wannier functions under {\it external} pressure, where the spread increases when the volume shrinks\cite{jmt_wannier,jmt_mno}.
When rescaling $V_{Wannier}$ with the unit-cell volume $V_0=a^3$, see \fref{Rfig}(e), we find the relative differences 
between the pairs with fixed precious metal ``M'' to more than halve.
Hence, for the Bi $\leftrightarrow$ Sb substitution, the intuitive correspondence between the size of ($p$-)orbitals that mediate bonding and the resulting lattice constant
holds to a large extent.

\floatsetup[figure]{style=plain,subcapbesideposition=bottom}
\floatsetup[figure]{style=plain,subcapbesideposition=top}

Finally, we discuss the extent of the Ce-$4f$ orbitals. The differences in $\overline{R}_f$ between the four compounds
may seem minute on the scale of \fref{Rfig}(a,b,c), yet their trends are revealing, see \fref{Rfig}(f):
Indeed, while for a given precious metal, a swapping of the pnictogen does not change the size of the $4f$-shell by much, replacing
the precious metal Pt with Pd significantly decreases $\overline{R}_f$ despite leaving the unit-cell volume inert.
This suggest that---for the chosen compounds---the impact of (i) the unit-cell volume and (ii) the {\it relative} extent of the $4f$-orbitals can be approximately disentangled, as discussed in the next section.

\subsection{Isoelectronic tuning: Summary \& Perspective}

\begin{figure}[!t]
  \begin{center}
{\scalebox{1.}{\includegraphics[clip=true,trim= 70 11 90 0, angle=0,width=.9\textwidth]{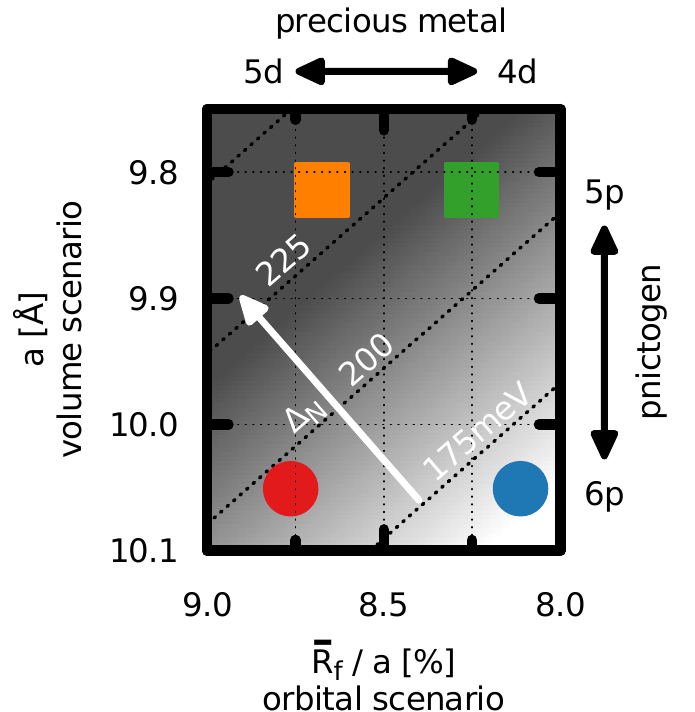}}} 
      \caption{{\bf Control parameters.}   The figure shows a (linearly interpolated) intensity map of the charge gap $\Delta_N$
			of \cbp\ \red{\ding{108}}, \cbpd\ \blue{\ding{108}}, \csp\ \yellow{\ding{110}}, and \cspd\ \green{\ding{110}}
			as a function of the lattice constant $a$ and the relative extent $\overline{R}_f/a$  of  the Ce-$4f$ orbitals.
			The substitution of the precious metal Pt$\leftrightarrow$Pd (horizontal arrow) only changes the size of the orbitals,
			while the pnictogen replacement Bi$\leftrightarrow$Sb (vertical arrow) dominantly
			engages the unit-cell volume via the lattice constant $a$. 
}      \label{scheme_new}
      \end{center}
\end{figure}

\fref{scheme_new} visualizes the central result of this paper: It
characterizes the charge gap of our four compounds as a function of their lattice constant $a$ and the relative extent $\overline{R}_f/a$ of their $f$-orbitals:
$\Delta_N=\Delta_N(a,\overline{R}_f/a)$.
Quite intuitively, $\Delta_N$ increases with shrinking volume and larger relative orbital extent, as both lead to larger orbital overlaps/hybridizations.
Most importantly, from this representation it is apparent, that the isoelectronic pnictogen replacement primarily acts on the lattice constant, while the relative
extent of the $f$-shell is well-neigh untouched.
The isoelectronic tuning mediated by an exchange of the precious metal, on the other hand, leaves the volume fixed, but 
expands or shrinks the $f$-orbitals' reach relative to the surrounding atoms.

We have thus identified a viable choice of essentially {\it independent} variables that connect our four materials in parameter space:
\fref{scheme_new} mirrors \fref{scheme} in a one-to-one fashion (if rotated by 40\deg to the right).

Let us put these findings into perspective:
It is a well-known fact that the Kondo coupling in $f$-electron materials can be dramatically affected by (i) pressure or by (ii) isoelectronic substitutions 
with atoms of varying size\cite{PSSB:PSSB201300005}.
Here, we demonstrated that these two routes can be {\it linearly independent}: Besides the effect of exercising chemical pressure, the change in the principle quantum number $n$
of substituted atoms---here Pt: $n=5$, Pd: $n=4$---may largely change the charge gap while having little or no effect on cohesive properties.
The isoelectronic tuning scenario that we propose extents the empirical volume picture (valid in binary compounds)
to include orbital effects at constant volume.
In this regard, the Ce$_3$A$_4$M$_3$ system is an ideal case, in that 
the changes in the lattice constant induced by different ``A'' atoms and the relative orbital extent of ``M''-atoms form an essentially orthogonal coordinate system.
In general, volume and $\overline{R}_f/a$ will provide linear independent yet skewed coordinates.
Near orthogonality of the control parameters $\overline{R}_f/a$ and the volume
 are likely realized for replacements of atoms---like Pt$\leftrightarrow$Pd---that have comparable
electronegativities and (empirical, van der Waals, or metallic) atomic radii (see below for a ``non-orthogonal'' example).
It also stands to reason that an isovolume tuning of inter-atomic hybridizations, and hence the Kondo coupling, 
is more prevalent in complex materials---ternary compounds and above---and
for substitutions of atoms that contribute only subleadingly to the charge gap/bonding (as is the case for the precious metal ``M'' in Ce$_3$A$_4$M$_3$, cf.\ \fref{scalehyb} and the discussion above).

The proposed orbital scenario will be relevant in other heavy fermion systems.
A putative example is the Ce(Ni,Pd)Sn system: CePdSn is an antiferromagnetic Kondo metal with a N{\'e}el temperature of about 7K\cite{KASAYA1988278,PhysRevB.40.2414,IGA1993419}, while CeNiSn
is a paramagnetic anisotropic Kondo insulator\cite{PhysRevB.41.9607,PhysRevLett.69.490}.
In the standard Doniach phase diagram\cite{DONIACH1977231}, CePdSn falls into the regime dominated by the Ruderman–Kittel–Kasuya–Yosida interaction, whereas CeNiSn realizes a larger Kondo coupling $J_K$ that 
impedes long-range magnetic order. This interpretation is congruent 
with the ``volume scenario'': CeNiSn has a unit-cell volume (263.6\AA$^3$) smaller by 6\% than CePdSn (281.7\AA$^3$)\cite{IGA1993419}
arguably leading to larger hybridizations and, hence, bigger $J_K$.
However, it has been highlighted\cite{PSSB:PSSB201300005} that there is a significant difference between 
the isoelectronic substitution series, Ce(Pd$_{1-x}$Ni$_x$)Sn, that connects the two compounds, and CePdSn under hydrostatic pressure.
Indeed, while the N{\'e}el temperature is fully suppressed for $x\geq 0.75$\cite{doi:10.1143/JPSJ.60.2542}, the pressure needed to achieve the same is at least 6GPa\cite{IGA1993419}, which---using the bulk modulus 55.6GPa of CeNiSn\cite{PhysRevB.60.14537}---corresponds to a volume reduction of more than 10\%:
an {\it external} compression of almost twice as much is required to equal the effect of {\it chemical} pressure.
With the ``orbital scenario'' proposed above, this large quantitative discrepancy is readily explained:
As suggested by the metallic radii $R_M$ of Pd and Ni\cite{Greenwood1997} (both are indicated in \fref{Rfig}(a)), the $3d$-orbitals of Ni 
will be smaller than the $4d$ of Pd.
We therefore hypothesize that for growing $x$ in CeNi$_{x}$Pd$_{1-x}$Sn the shrinking of the unit-cell volume
and the decrease of the orbital extent conspire to jointly reduce the Kondo coupling---driving the system faster towards smaller $J_K$-coupling in the Doniach phase diagram.
The scenario might also be of relevance in the comparison of CeRu$_4$Sn$_6$\cite{PhysRevB.46.4250} under external\cite{Sengupta_2012,Zhang_2018} and chemical (Sn$\rightarrow$Ge)\cite{Zhang_2018} pressure.
In fact, it will be interesting to gain systematic insights into larger groups of materials to assess the described physics. 
Indeed, we believe that $\overline{R}_f/a$ could be a useful quantity in databases for heavy-fermion material properties\cite{PhysRevMaterials.1.033802,Hafiz2018}.
Its approximation $R_M/a$, using tabulated metallic radii $R_M$, could even be used as descriptor in high-throughput surveys.%
\footnote{For non-cubic compounds, such as CeNiSn and CeRu$_4$Sn$_6$, the lattice parameter $a$ might have to be replaced with relevant nearest neighbour distances.}

\subsection{Anticipated many-body renormalizations}\label{sec:hyb}  

Finally, we transition to a more many-body oriented language.
In the spirit of the Anderson impurity model, the embedding of the correlated Ce-$4f$ orbitals into the solid is often described
by a so-called hybridization function $\Delta(\omega)=\omega+\mu-H^{loc}_{ff}-(G_{loc}^{-1})_{ff}$. Here, $\mu$ is the chemical potential, and 
$H^{loc}$ and $G_{loc}$ are the local projections\cite{PhysRevB.81.195107} of the Hamiltonian and the Greensfunction, respectively, while the subscripts $f$ denotes the evaluation within the $f$-orbital subspace.
Indeed, $\Delta(\omega)$ is the back-bone of dynamical mean-field theory (DMFT)\cite{bible}, where it is promoted to a self-consistent Weiss field. 
However, already on the level of DFT, the hybridization function can divulge important physical insight, in particular regarding {\it trends}
among material families\cite{PhysRevMaterials.1.033802,Hafiz2018}.
Moreover, $\Delta(\omega)$ can be used to classify insulators\cite{jmt_CBP_arxiv}, e.g., distinguishing between Mott and Kondo insulators.

\begin{figure*}[!t]
  \begin{center}
\sidesubfloat[]{\scalebox{1.}{\includegraphics[clip=true,trim= 0 40 0 20, angle=0,angle=0,width=.45\textwidth]{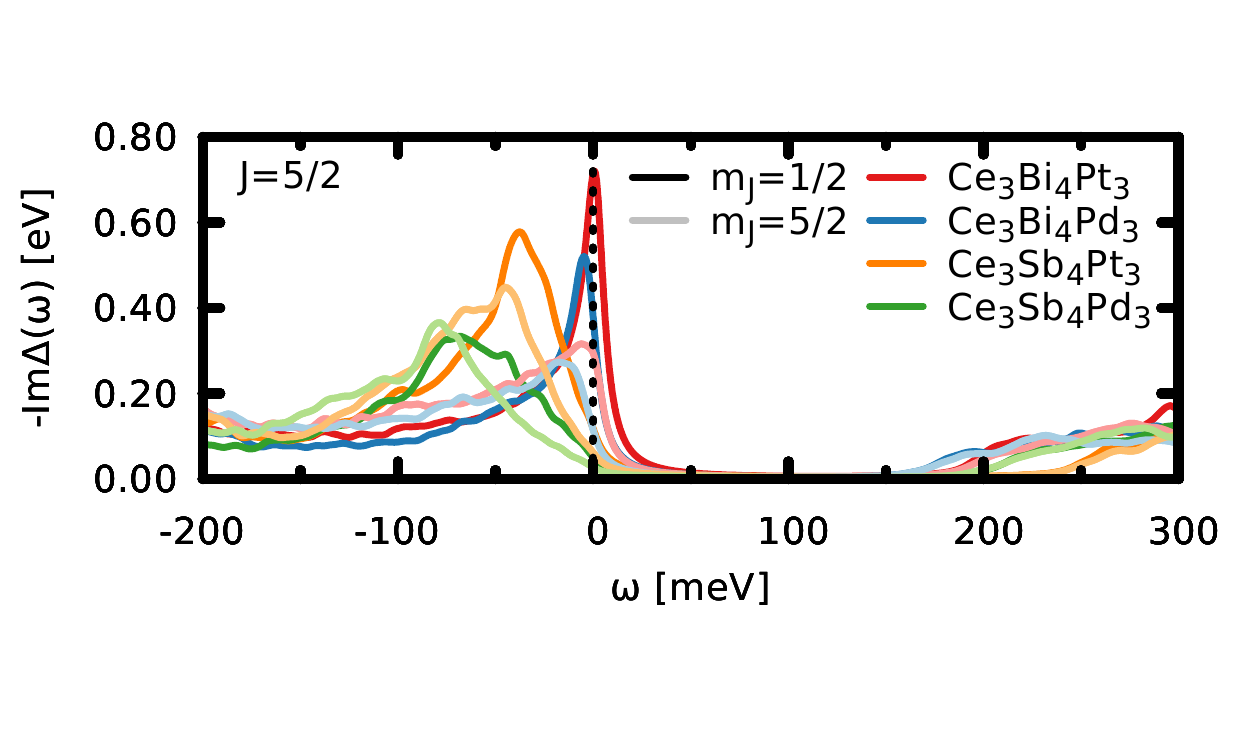}}} \quad%
\sidesubfloat[]{\scalebox{1.}{\includegraphics[clip=true,trim= 0 40 0 20, angle=0,angle=0,width=.45\textwidth]{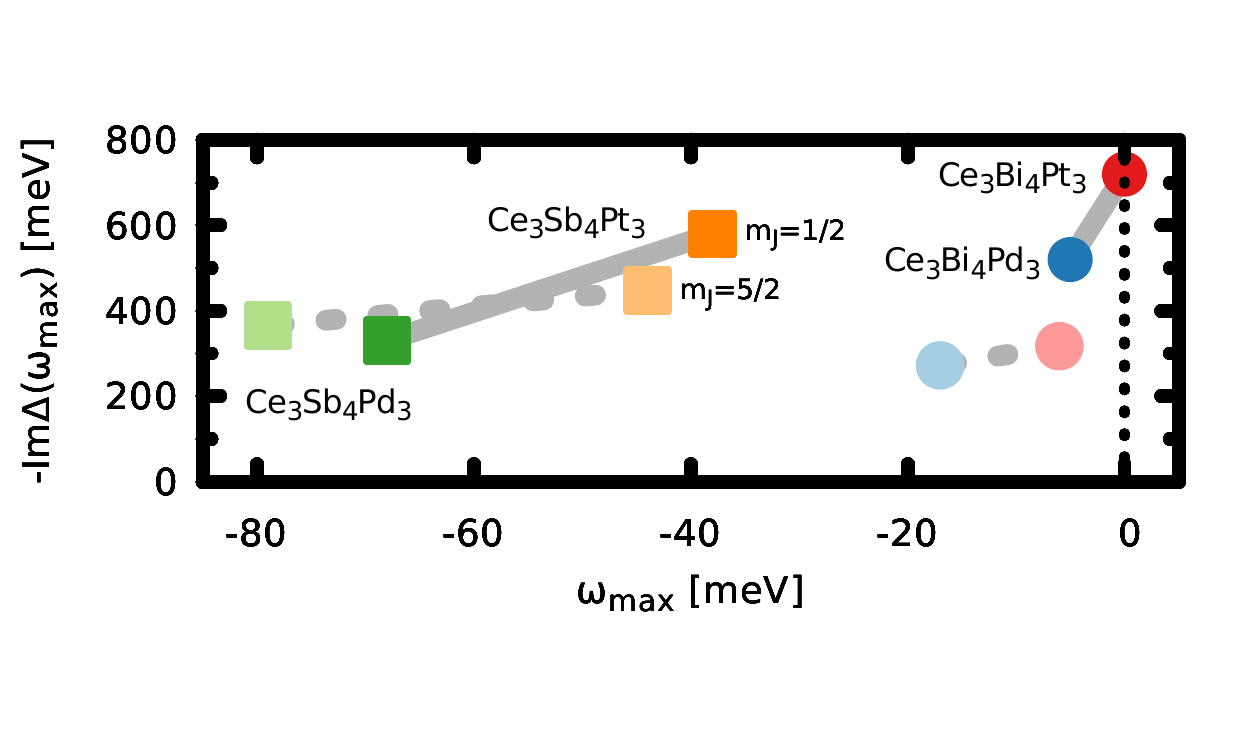}}} 
      \caption{{\bf Hybridization function $\Delta(\omega)$.} 
			Panel (a) shows the imaginary part of the hybridization function $\Delta(\omega)$ of all compounds, resolved into their $m_J$ components.
			Panel (b) indicates the trends in the maximal amplitude of the $m_J=1/2$ component of $\Delta(\omega)$ and the frequency $\omega_{max}$ where the latter occurs.
			The origin of energy corresponds to the valence-band maximum.}
      \label{dlt}
      \end{center}
\end{figure*}

For our four materials, \fref{dlt}(a) displays the imaginary parts of the DFT hybridization function,
 while panel (b) shows the corresponding positions $\omega_{max}$ and peak amplitudes $-\Im\Delta(\omega_{max})$.
In both cases, the origin of energy corresponds to the top of the valence bands, and 
we have limited the presentation to the $m_J=1/2$ and $m_J=5/2$ components of the $J=5/2$ multiplet (the $m_J=3/2$ components are the smallest).

Overall, the Bi-compounds exhibit peaks that are sharper and closer to the Fermi level, indicating that these materials
are closer to the flat-band limit. In the latter, an $f$-level hybridizes with a dispersionless conduction band, yielding
$-\Im\Delta(\omega)=\pi V^2\delta(\omega+\mu-\omega_{max})$, where $V_\svek{k}=V$ is the amplitude of the hybridization between the conduction band and the Ce-$4f$ orbitals and $\epsilon_\svek{k}=\omega_{max}$ the position of the
conduction band. For conduction-states $\epsilon_\svek{k}$ and couplings $V_\svek{k}$ with finite momentum-dependence, the 
hybridization strength is spread over larger energies. The latter is the case for the Sb-members of the Ce$_3$A$_4$M$_3$ family.
For a given pnictogen ``A'', the hybridization function is much more pronounced for Pt than for Pd, as could be anticipated from
the size of orbitals ($5d$ vs.\ $4d$) as discussed in the previous section.

Using the DFT hybridization function in conjunction with the experimental magnetic susceptibility, we can estimate the magnitude of 
many-body renormalizations, see \tref{tab1}.
We first assess the Kondo temperature from the spin degrees of freedom using the Kondo lattice expression, $T_K=4T^{max}_\chi/(2J+1)$\cite{PhysRevB.28.5255} with $J=5/2$ and $T^{max}_\chi$ from \fref{rhochi}.
Note that the estimate for $T_K$ of \cbpd\ is of the same order of magnitude of what has been deduced from the experimental specific heat\cite{PhysRevLett.118.246601}.
The Kondo temperature can also be determined from the charge degrees of freedom: $k_BT_K=-(\pi/4)\,\Im\Delta(\omega_{max})\times Z/(2J+1)$ \cite{hewson,PhysRevB.28.5255}.
With the $T_K$ from above, we then extract the effective mass enhancement $1/Z$.
For \cbp, we find $Z\approx 1/10$, in excellent agreement with recent realistic DMFT calculations\cite{jmt_CBP_arxiv}.
While for the other compounds, a quantitative theoretical assessment has to await the application of many-body electronic structure methods\cite{bible,Tomczak2017review}%
$^,$\footnote{
For the temperature dependence of $\Delta(\omega)$ in \cbp, see Ref.~\onlinecite{jmt_CBP_arxiv};
for spectral properties of \cbp\ and \cbpd, see Refs.~\onlinecite{jmt_CBP_arxiv,Cao2019}.
}, 
we believe the following simple estimates to reflect at least the {\it hierarchy} of effective masses in the Ce$_3$A$_4$M$_3$ family:  
The largely reduced Kondo temperature when substituting Pd for Pt accounts for \cbpd\ having the largest effective mass ($1/Z=66$), as well as for
\csp\ being less correlated than \cbp.
Naively applying the mass renormalization of \cbpd\ to its DFT band-gap yields 2.6meV, in good agreement with the experimental estimate of $1.8\pm0.5$meV\cite{Kushwaha2019}.
Doing the same for \csp, however, significantly overestimates the gap. We can surmise that the Sb-compounds are further from a description in terms of fully localized $f$-states,
which renders the Kondo lattice formula used in the determination of $T_K$ less reliable.

Given the trends in the band gaps, hybridization functions, and the Kondo temperature (that itself depends on the two former), we nonetheless gauge the
\cspd\ compound---that we propose experimentalists to synthesize---to exhibit mass enhancements of the order of 10.

\begin{table}
\begin{tabular}{c||c|c|c|c|c}
 & {exp} & &{theory} & \multicolumn{2}{c}{prediction}\\
\hline
material & $T^{max}_\chi$  & $T_K$ & $\left\langle -\Im\Delta(\omega_{max})\right\rangle_{m_J}$ & $m^*/m_{band}$& $\Delta$\\
 & [K] & [K] &  [meV] & $=1/Z$&[meV]\\
\hline 
\red{\cbp}  &  80 &   53 & 400 & 11 & 17\\
\blue{\cbpd} &  10 &    7 & 290 & 66 & 2.6\\
\yellow{\csp}  & 400 &  265 & 480 & 2.7 & 90\\
\green{\cspd}  & ? &  ? & 300 & ? & ?\\
\end{tabular}
\caption{Collection of experimental $T^{max}_\chi$ (see \fref{rhochi}), estimated Kondo temperatures $T_K=4T^{max}_\chi/(2J+1)$ with $J=5/2$, 
$m_J$-averaged peak magnitude of the hybridization function $-\Im\Delta(\omega_{max})$, and
anticipated mass renormalization $1/Z$ estimated via $k_BT_K=-(\pi/4)\,\Im\Delta(\omega_{max})\times Z/(2J+1)$;
the gap $\Delta$ is predicted via $Z\times\Delta_{DFT}$.
}
\label{tab1}
\end{table}

\section{Conclusions.}

We have shown the suppression of coherence and energy scales in isoelectronic and isovolume Ce$_3$Bi$_4$(Pt$_{1-x}$Pd$_x$)$_3$---previously attributed to changes of the spin-orbit coupling\cite{PhysRevLett.118.246601}---to derive from a reduced Kondo coupling.
The latter is tunable---even at constant volume---through the principle quantum number of orbitals that host conduction electrons.
We have identified the relative extent, $\overline{R}_f/a$, of the lanthanide's $f$-orbitals as (indirect) control parameter of the Kondo coupling
and demonstrated it to be {\it linearly independent} from changes mediated by the unit-cell volume $V_0$:
In the Ce$_3$A$_4$M$_3$ system, the precious metal Pt$\leftrightarrow$Pd (pnictide Bi$\leftrightarrow$Sb) substitution only
changes the $f$-orbitals' hybridizations through $\overline{R}_f/a$ ($V_0$).
Our ``orbital scenario'' generalizes the empirical ``volume scenario'' (applicable to binary compounds)
to isoelectronic tuning in complex heavy-fermion materials.
In particular, the two parameter dependency $J_K=J_K(a,\overline{R}_f/a)$
explains why there can be notable differences between external and chemical pressure, cf.\ the Ce(Ni,Pd)Sn system.

Since \cbpd\ has a very small coherence scale, its ground-state remains controversial\cite{PhysRevLett.118.246601,Kushwaha2019}.
We therefore propose to synthesise
the new material \cspd. We predict it to have a unit-cell volume very close to that of \csp, and a mass enhancement of the order of 10.
Anticipating larger gaps in  Ce$_3$Sb$_4$(Pt$_{1-x}$Pd$_x$)$_3$, the series provides an attractive testing ground for 
the proposed ``orbital scenario'' of
isoelectronic (and isovolume) tuning of energy and coherence scales.

\section{Acknowledgements}

The author is indebted to Juan Fernandez Afonso.
Exchanges with
S.\ Dzsaber, G.\ Eguchi, K.\ Held, J.\ Kune\v{s}, S.\ Paschen, and Q.\ Si are acknowledged.
This work has been supported by the Austrian Science Fund (FWF)
through project ``LinReTraCe'' P~30213-N36.
Some calculations were performed on the Vienna Scientific Cluster (VSC).


\begin{thebibliography}{75}%
\makeatletter
\providecommand \@ifxundefined [1]{%
 \@ifx{#1\undefined}
}%
\providecommand \@ifnum [1]{%
 \ifnum #1\expandafter \@firstoftwo
 \else \expandafter \@secondoftwo
 \fi
}%
\providecommand \@ifx [1]{%
 \ifx #1\expandafter \@firstoftwo
 \else \expandafter \@secondoftwo
 \fi
}%
\providecommand \natexlab [1]{#1}%
\providecommand \enquote  [1]{``#1''}%
\providecommand \bibnamefont  [1]{#1}%
\providecommand \bibfnamefont [1]{#1}%
\providecommand \citenamefont [1]{#1}%
\providecommand \href@noop [0]{\@secondoftwo}%
\providecommand \href [0]{\begingroup \@sanitize@url \@href}%
\providecommand \@href[1]{\@@startlink{#1}\@@href}%
\providecommand \@@href[1]{\endgroup#1\@@endlink}%
\providecommand \@sanitize@url [0]{\catcode `\\12\catcode `\$12\catcode
  `\&12\catcode `\#12\catcode `\^12\catcode `\_12\catcode `\%12\relax}%
\providecommand \@@startlink[1]{}%
\providecommand \@@endlink[0]{}%
\providecommand \url  [0]{\begingroup\@sanitize@url \@url }%
\providecommand \@url [1]{\endgroup\@href {#1}{\urlprefix }}%
\providecommand \urlprefix  [0]{URL }%
\providecommand \Eprint [0]{\href }%
\providecommand \doibase [0]{http://dx.doi.org/}%
\providecommand \selectlanguage [0]{\@gobble}%
\providecommand \bibinfo  [0]{\@secondoftwo}%
\providecommand \bibfield  [0]{\@secondoftwo}%
\providecommand \translation [1]{[#1]}%
\providecommand \BibitemOpen [0]{}%
\providecommand \bibitemStop [0]{}%
\providecommand \bibitemNoStop [0]{.\EOS\space}%
\providecommand \EOS [0]{\spacefactor3000\relax}%
\providecommand \BibitemShut  [1]{\csname bibitem#1\endcsname}%
\let\auto@bib@innerbib\@empty
\bibitem [{\citenamefont {Gegenwart}\ \emph {et~al.}(2008)\citenamefont
  {Gegenwart}, \citenamefont {Si},\ and\ \citenamefont
  {Steglich}}]{Gegenwart2008}%
  \BibitemOpen
  \bibfield  {author} {\bibinfo {author} {\bibfnamefont {P.}~\bibnamefont
  {Gegenwart}}, \bibinfo {author} {\bibfnamefont {Q.}~\bibnamefont {Si}}, \
  and\ \bibinfo {author} {\bibfnamefont {F.}~\bibnamefont {Steglich}},\ }\href
  {\doibase 10.1038/nphys892} {\bibfield  {journal} {\bibinfo  {journal} {Nat
  Phys}\ }\textbf {\bibinfo {volume} {4}},\ \bibinfo {pages} {186} (\bibinfo
  {year} {2008})}\BibitemShut {NoStop}%
\bibitem [{\citenamefont {Si}\ and\ \citenamefont
  {Paschen}(2013)}]{PSSB:PSSB201300005}%
  \BibitemOpen
  \bibfield  {author} {\bibinfo {author} {\bibfnamefont {Q.}~\bibnamefont
  {Si}}\ and\ \bibinfo {author} {\bibfnamefont {S.}~\bibnamefont {Paschen}},\
  }\href {\doibase 10.1002/pssb.201300005} {\bibfield  {journal} {\bibinfo
  {journal} {physica status solidi (b)}\ }\textbf {\bibinfo {volume} {250}},\
  \bibinfo {pages} {425} (\bibinfo {year} {2013})}\BibitemShut {NoStop}%
\bibitem [{\citenamefont {Wirth}\ and\ \citenamefont
  {Steglich}(2016)}]{Wirth2016}%
  \BibitemOpen
  \bibfield  {author} {\bibinfo {author} {\bibfnamefont {S.}~\bibnamefont
  {Wirth}}\ and\ \bibinfo {author} {\bibfnamefont {F.}~\bibnamefont
  {Steglich}},\ }\href {http://dx.doi.org/10.1038/natrevmats.2016.51}
  {\bibfield  {journal} {\bibinfo  {journal} {Nat. Rev. Materials}\ }\textbf
  {\bibinfo {volume} {1}},\ \bibinfo {pages} {16051 EP } (\bibinfo {year}
  {2016})}\BibitemShut {NoStop}%
\bibitem [{\citenamefont {Tomczak}(2018)}]{NGCS}%
  \BibitemOpen
  \bibfield  {author} {\bibinfo {author} {\bibfnamefont {J.~M.}\ \bibnamefont
  {Tomczak}},\ }\href {http://stacks.iop.org/0953-8984/30/i=18/a=183001}
  {\bibfield  {journal} {\bibinfo  {journal} {J. Phys.: Condens. Matter
  (Topical Review)}\ }\textbf {\bibinfo {volume} {30}},\ \bibinfo {pages}
  {183001} (\bibinfo {year} {2018})}\BibitemShut {NoStop}%
\bibitem [{\citenamefont {Doniach}(1977)}]{DONIACH1977231}%
  \BibitemOpen
  \bibfield  {author} {\bibinfo {author} {\bibfnamefont {S.}~\bibnamefont
  {Doniach}},\ }\href {\doibase http://dx.doi.org/10.1016/0378-4363(77)90190-5}
  {\bibfield  {journal} {\bibinfo  {journal} {Physica B+C}\ }\textbf {\bibinfo
  {volume} {91}},\ \bibinfo {pages} {231 } (\bibinfo {year}
  {1977})}\BibitemShut {NoStop}%
\bibitem [{\citenamefont {Herper}\ \emph {et~al.}(2017)\citenamefont {Herper},
  \citenamefont {Ahmed}, \citenamefont {Wills}, \citenamefont {Di~Marco},
  \citenamefont {Bj\"orkman}, \citenamefont {Iu\ifmmode~\mbox{\c{s}}\else
  \c{s}\fi{}an}, \citenamefont {Balatsky},\ and\ \citenamefont
  {Eriksson}}]{PhysRevMaterials.1.033802}%
  \BibitemOpen
  \bibfield  {author} {\bibinfo {author} {\bibfnamefont {H.~C.}\ \bibnamefont
  {Herper}}, \bibinfo {author} {\bibfnamefont {T.}~\bibnamefont {Ahmed}},
  \bibinfo {author} {\bibfnamefont {J.~M.}\ \bibnamefont {Wills}}, \bibinfo
  {author} {\bibfnamefont {I.}~\bibnamefont {Di~Marco}}, \bibinfo {author}
  {\bibfnamefont {T.}~\bibnamefont {Bj\"orkman}}, \bibinfo {author}
  {\bibfnamefont {D.}~\bibnamefont {Iu\ifmmode~\mbox{\c{s}}\else
  \c{s}\fi{}an}}, \bibinfo {author} {\bibfnamefont {A.~V.}\ \bibnamefont
  {Balatsky}}, \ and\ \bibinfo {author} {\bibfnamefont {O.}~\bibnamefont
  {Eriksson}},\ }\href {\doibase 10.1103/PhysRevMaterials.1.033802} {\bibfield
  {journal} {\bibinfo  {journal} {Phys. Rev. Materials}\ }\textbf {\bibinfo
  {volume} {1}},\ \bibinfo {pages} {033802} (\bibinfo {year}
  {2017})}\BibitemShut {NoStop}%
\bibitem [{\citenamefont {Hundley}\ \emph {et~al.}(1990)\citenamefont
  {Hundley}, \citenamefont {Canfield}, \citenamefont {Thompson}, \citenamefont
  {Fisk},\ and\ \citenamefont {Lawrence}}]{PhysRevB.42.6842}%
  \BibitemOpen
  \bibfield  {author} {\bibinfo {author} {\bibfnamefont {M.~F.}\ \bibnamefont
  {Hundley}}, \bibinfo {author} {\bibfnamefont {P.~C.}\ \bibnamefont
  {Canfield}}, \bibinfo {author} {\bibfnamefont {J.~D.}\ \bibnamefont
  {Thompson}}, \bibinfo {author} {\bibfnamefont {Z.}~\bibnamefont {Fisk}}, \
  and\ \bibinfo {author} {\bibfnamefont {J.~M.}\ \bibnamefont {Lawrence}},\
  }\href {\doibase 10.1103/PhysRevB.42.6842} {\bibfield  {journal} {\bibinfo
  {journal} {Phys. Rev. B}\ }\textbf {\bibinfo {volume} {42}},\ \bibinfo
  {pages} {6842} (\bibinfo {year} {1990})}\BibitemShut {NoStop}%
\bibitem [{\citenamefont {Kwei}\ \emph {et~al.}(1992)\citenamefont {Kwei},
  \citenamefont {Lawrence}, \citenamefont {Canfield}, \citenamefont
  {Beyermann}, \citenamefont {Thompson}, \citenamefont {Fisk}, \citenamefont
  {Lawson},\ and\ \citenamefont {Goldstone}}]{PhysRevB.46.8067}%
  \BibitemOpen
  \bibfield  {author} {\bibinfo {author} {\bibfnamefont {G.~H.}\ \bibnamefont
  {Kwei}}, \bibinfo {author} {\bibfnamefont {J.~M.}\ \bibnamefont {Lawrence}},
  \bibinfo {author} {\bibfnamefont {P.~C.}\ \bibnamefont {Canfield}}, \bibinfo
  {author} {\bibfnamefont {W.~P.}\ \bibnamefont {Beyermann}}, \bibinfo {author}
  {\bibfnamefont {J.~D.}\ \bibnamefont {Thompson}}, \bibinfo {author}
  {\bibfnamefont {Z.}~\bibnamefont {Fisk}}, \bibinfo {author} {\bibfnamefont
  {A.~C.}\ \bibnamefont {Lawson}}, \ and\ \bibinfo {author} {\bibfnamefont
  {J.~A.}\ \bibnamefont {Goldstone}},\ }\href {\doibase
  10.1103/PhysRevB.46.8067} {\bibfield  {journal} {\bibinfo  {journal} {Phys.
  Rev. B}\ }\textbf {\bibinfo {volume} {46}},\ \bibinfo {pages} {8067}
  (\bibinfo {year} {1992})}\BibitemShut {NoStop}%
\bibitem [{\citenamefont {Hermes}\ \emph {et~al.}(2008)\citenamefont {Hermes},
  \citenamefont {Linsinger}, \citenamefont {Mishra},\ and\ \citenamefont
  {P{\"o}ttgen}}]{Hermes2008}%
  \BibitemOpen
  \bibfield  {author} {\bibinfo {author} {\bibfnamefont {W.}~\bibnamefont
  {Hermes}}, \bibinfo {author} {\bibfnamefont {S.}~\bibnamefont {Linsinger}},
  \bibinfo {author} {\bibfnamefont {R.}~\bibnamefont {Mishra}}, \ and\ \bibinfo
  {author} {\bibfnamefont {R.}~\bibnamefont {P{\"o}ttgen}},\ }\href {\doibase
  10.1007/s00706-008-0914-4} {\bibfield  {journal} {\bibinfo  {journal}
  {Monatshefte f{\"u}r Chemie - Chemical Monthly}\ }\textbf {\bibinfo {volume}
  {139}},\ \bibinfo {pages} {1143} (\bibinfo {year} {2008})}\BibitemShut
  {NoStop}%
\bibitem [{\citenamefont {Fisk}\ \emph {et~al.}(1995)\citenamefont {Fisk},
  \citenamefont {Sarrao}, \citenamefont {Thompson}, \citenamefont {Mandrus},
  \citenamefont {Hundley}, \citenamefont {Miglori}, \citenamefont {Bucher},
  \citenamefont {Schlesinger}, \citenamefont {Aeppli}, \citenamefont {Bucher},
  \citenamefont {DiTusa}, \citenamefont {Oglesby}, \citenamefont {Ott},
  \citenamefont {Canfield},\ and\ \citenamefont {Brown}}]{FISK1995798}%
  \BibitemOpen
  \bibfield  {author} {\bibinfo {author} {\bibfnamefont {Z.}~\bibnamefont
  {Fisk}}, \bibinfo {author} {\bibfnamefont {J.}~\bibnamefont {Sarrao}},
  \bibinfo {author} {\bibfnamefont {J.}~\bibnamefont {Thompson}}, \bibinfo
  {author} {\bibfnamefont {D.}~\bibnamefont {Mandrus}}, \bibinfo {author}
  {\bibfnamefont {M.}~\bibnamefont {Hundley}}, \bibinfo {author} {\bibfnamefont
  {A.}~\bibnamefont {Miglori}}, \bibinfo {author} {\bibfnamefont
  {B.}~\bibnamefont {Bucher}}, \bibinfo {author} {\bibfnamefont
  {Z.}~\bibnamefont {Schlesinger}}, \bibinfo {author} {\bibfnamefont
  {G.}~\bibnamefont {Aeppli}}, \bibinfo {author} {\bibfnamefont
  {E.}~\bibnamefont {Bucher}}, \bibinfo {author} {\bibfnamefont
  {J.}~\bibnamefont {DiTusa}}, \bibinfo {author} {\bibfnamefont
  {C.}~\bibnamefont {Oglesby}}, \bibinfo {author} {\bibfnamefont {H.-R.}\
  \bibnamefont {Ott}}, \bibinfo {author} {\bibfnamefont {P.}~\bibnamefont
  {Canfield}}, \ and\ \bibinfo {author} {\bibfnamefont {S.}~\bibnamefont
  {Brown}},\ }\href {\doibase http://dx.doi.org/10.1016/0921-4526(94)00588-M}
  {\bibfield  {journal} {\bibinfo  {journal} {Physica B:}\ }\textbf {\bibinfo
  {volume} {206}},\ \bibinfo {pages} {798 } (\bibinfo {year}
  {1995})}\BibitemShut {NoStop}%
\bibitem [{\citenamefont {Riseborough}(2000)}]{Riseborough2000}%
  \BibitemOpen
  \bibfield  {author} {\bibinfo {author} {\bibfnamefont {P.~S.}\ \bibnamefont
  {Riseborough}},\ }\href {\doibase 10.1080/000187300243345} {\bibfield
  {journal} {\bibinfo  {journal} {Advances in Physics}\ }\textbf {\bibinfo
  {volume} {49}},\ \bibinfo {pages} {257} (\bibinfo {year} {2000})}\BibitemShut
  {NoStop}%
\bibitem [{\citenamefont {Dzsaber}\ \emph {et~al.}(2017)\citenamefont
  {Dzsaber}, \citenamefont {Prochaska}, \citenamefont {Sidorenko},
  \citenamefont {Eguchi}, \citenamefont {Svagera}, \citenamefont {Waas},
  \citenamefont {Prokofiev}, \citenamefont {Si},\ and\ \citenamefont
  {Paschen}}]{PhysRevLett.118.246601}%
  \BibitemOpen
  \bibfield  {author} {\bibinfo {author} {\bibfnamefont {S.}~\bibnamefont
  {Dzsaber}}, \bibinfo {author} {\bibfnamefont {L.}~\bibnamefont {Prochaska}},
  \bibinfo {author} {\bibfnamefont {A.}~\bibnamefont {Sidorenko}}, \bibinfo
  {author} {\bibfnamefont {G.}~\bibnamefont {Eguchi}}, \bibinfo {author}
  {\bibfnamefont {R.}~\bibnamefont {Svagera}}, \bibinfo {author} {\bibfnamefont
  {M.}~\bibnamefont {Waas}}, \bibinfo {author} {\bibfnamefont {A.}~\bibnamefont
  {Prokofiev}}, \bibinfo {author} {\bibfnamefont {Q.}~\bibnamefont {Si}}, \
  and\ \bibinfo {author} {\bibfnamefont {S.}~\bibnamefont {Paschen}},\ }\href
  {\doibase 10.1103/PhysRevLett.118.246601} {\bibfield  {journal} {\bibinfo
  {journal} {Phys. Rev. Lett.}\ }\textbf {\bibinfo {volume} {118}},\ \bibinfo
  {pages} {246601} (\bibinfo {year} {2017})}\BibitemShut {NoStop}%
\bibitem [{\citenamefont {{Dzsaber}}\ \emph {et~al.}(2018)\citenamefont
  {{Dzsaber}}, \citenamefont {{Yan}}, \citenamefont {{Eguchi}}, \citenamefont
  {{Prokofiev}}, \citenamefont {{Shiroka}}, \citenamefont {{Blaha}},
  \citenamefont {{Rubel}}, \citenamefont {{Grefe}}, \citenamefont {{Lai}},
  \citenamefont {{Si}},\ and\ \citenamefont {{Paschen}}}]{Dzsaber2018arxiv}%
  \BibitemOpen
  \bibfield  {author} {\bibinfo {author} {\bibfnamefont {S.}~\bibnamefont
  {{Dzsaber}}}, \bibinfo {author} {\bibfnamefont {X.}~\bibnamefont {{Yan}}},
  \bibinfo {author} {\bibfnamefont {G.}~\bibnamefont {{Eguchi}}}, \bibinfo
  {author} {\bibfnamefont {A.}~\bibnamefont {{Prokofiev}}}, \bibinfo {author}
  {\bibfnamefont {T.}~\bibnamefont {{Shiroka}}}, \bibinfo {author}
  {\bibfnamefont {P.}~\bibnamefont {{Blaha}}}, \bibinfo {author} {\bibfnamefont
  {O.}~\bibnamefont {{Rubel}}}, \bibinfo {author} {\bibfnamefont {S.~E.}\
  \bibnamefont {{Grefe}}}, \bibinfo {author} {\bibfnamefont {H.-H.}\
  \bibnamefont {{Lai}}}, \bibinfo {author} {\bibfnamefont {Q.}~\bibnamefont
  {{Si}}}, \ and\ \bibinfo {author} {\bibfnamefont {S.}~\bibnamefont
  {{Paschen}}},\ }\href@noop {} {\bibfield  {journal} {\bibinfo  {journal}
  {arXiv e-prints}\ ,\ \bibinfo {eid} {arXiv:1811.02819}} (\bibinfo {year}
  {2018})},\ \Eprint {http://arxiv.org/abs/1811.02819} {arXiv:1811.02819
  [cond-mat.str-el]} \BibitemShut {NoStop}%
\bibitem [{\citenamefont {{Tomczak}}(2019)}]{jmt_CBP_arxiv}%
  \BibitemOpen
  \bibfield  {author} {\bibinfo {author} {\bibfnamefont {J.~M.}\ \bibnamefont
  {{Tomczak}}},\ }\href@noop {} {\bibfield  {journal} {\bibinfo  {journal}
  {arXiv e-prints}\ ,\ \bibinfo {eid} {arXiv:1904.01346}} (\bibinfo {year}
  {2019})},\ \Eprint {http://arxiv.org/abs/1904.01346} {arXiv:1904.01346
  [cond-mat.str-el]} \BibitemShut {NoStop}%
\bibitem [{\citenamefont {{Dzsaber}}\ \emph {et~al.}(2019)\citenamefont
  {{Dzsaber}}, \citenamefont {{Zocco}}, \citenamefont {{McCollam}},
  \citenamefont {{Weickert}}, \citenamefont {{McDonald}}, \citenamefont
  {{Taupin}}, \citenamefont {{Yan}}, \citenamefont {{Prokofiev}}, \citenamefont
  {{Tang}}, \citenamefont {{Vlaar}}, \citenamefont {{Stritzinger}},
  \citenamefont {{Jaime}}, \citenamefont {{Si}},\ and\ \citenamefont
  {{Paschen}}}]{Dzsaber2019arxiv}%
  \BibitemOpen
  \bibfield  {author} {\bibinfo {author} {\bibfnamefont {S.}~\bibnamefont
  {{Dzsaber}}}, \bibinfo {author} {\bibfnamefont {D.~A.}\ \bibnamefont
  {{Zocco}}}, \bibinfo {author} {\bibfnamefont {A.}~\bibnamefont {{McCollam}}},
  \bibinfo {author} {\bibfnamefont {F.}~\bibnamefont {{Weickert}}}, \bibinfo
  {author} {\bibfnamefont {R.}~\bibnamefont {{McDonald}}}, \bibinfo {author}
  {\bibfnamefont {M.}~\bibnamefont {{Taupin}}}, \bibinfo {author}
  {\bibfnamefont {X.}~\bibnamefont {{Yan}}}, \bibinfo {author} {\bibfnamefont
  {A.}~\bibnamefont {{Prokofiev}}}, \bibinfo {author} {\bibfnamefont
  {L.~M.~K.}\ \bibnamefont {{Tang}}}, \bibinfo {author} {\bibfnamefont
  {B.}~\bibnamefont {{Vlaar}}}, \bibinfo {author} {\bibfnamefont
  {L.}~\bibnamefont {{Stritzinger}}}, \bibinfo {author} {\bibfnamefont
  {M.}~\bibnamefont {{Jaime}}}, \bibinfo {author} {\bibfnamefont
  {Q.}~\bibnamefont {{Si}}}, \ and\ \bibinfo {author} {\bibfnamefont
  {S.}~\bibnamefont {{Paschen}}},\ }\href@noop {} {\bibfield  {journal}
  {\bibinfo  {journal} {arXiv e-prints}\ ,\ \bibinfo {eid} {arXiv:1906.01182}}
  (\bibinfo {year} {2019})},\ \Eprint {http://arxiv.org/abs/1906.01182}
  {arXiv:1906.01182 [cond-mat.str-el]} \BibitemShut {NoStop}%
\bibitem [{\citenamefont {{Cao}}\ \emph {et~al.}(2019)\citenamefont {{Cao}},
  \citenamefont {{Zhi}},\ and\ \citenamefont {{Zhu}}}]{Cao2019}%
  \BibitemOpen
  \bibfield  {author} {\bibinfo {author} {\bibfnamefont {C.}~\bibnamefont
  {{Cao}}}, \bibinfo {author} {\bibfnamefont {G.-X.}\ \bibnamefont {{Zhi}}}, \
  and\ \bibinfo {author} {\bibfnamefont {J.-X.}\ \bibnamefont {{Zhu}}},\
  }\href@noop {} {\bibfield  {journal} {\bibinfo  {journal} {arXiv e-prints}\
  ,\ \bibinfo {eid} {arXiv:1904.00675}} (\bibinfo {year} {2019})},\ \Eprint
  {http://arxiv.org/abs/1904.00675} {arXiv:1904.00675 [cond-mat.str-el]}
  \BibitemShut {NoStop}%
\bibitem [{\citenamefont {{Kushwaha}}\ \emph {et~al.}(2019)\citenamefont
  {{Kushwaha}}, \citenamefont {{Chan}}, \citenamefont {{Park}}, \citenamefont
  {{Thomas}}, \citenamefont {{Bauer}}, \citenamefont {{Thompson}},
  \citenamefont {{Ronning}}, \citenamefont {{Rosa}},\ and\ \citenamefont
  {{Harrison}}}]{Kushwaha2019}%
  \BibitemOpen
  \bibfield  {author} {\bibinfo {author} {\bibfnamefont {S.~K.}\ \bibnamefont
  {{Kushwaha}}}, \bibinfo {author} {\bibfnamefont {M.~K.}\ \bibnamefont
  {{Chan}}}, \bibinfo {author} {\bibfnamefont {J.}~\bibnamefont {{Park}}},
  \bibinfo {author} {\bibfnamefont {S.~M.}\ \bibnamefont {{Thomas}}}, \bibinfo
  {author} {\bibfnamefont {E.~D.}\ \bibnamefont {{Bauer}}}, \bibinfo {author}
  {\bibfnamefont {J.~D.}\ \bibnamefont {{Thompson}}}, \bibinfo {author}
  {\bibfnamefont {F.}~\bibnamefont {{Ronning}}}, \bibinfo {author}
  {\bibfnamefont {P.~F.~S.}\ \bibnamefont {{Rosa}}}, \ and\ \bibinfo {author}
  {\bibfnamefont {N.}~\bibnamefont {{Harrison}}},\ }\href@noop {} {\bibfield
  {journal} {\bibinfo  {journal} {arXiv e-prints}\ ,\ \bibinfo {eid}
  {arXiv:1906.01740}} (\bibinfo {year} {2019})},\ \Eprint
  {http://arxiv.org/abs/1906.01740} {arXiv:1906.01740 [cond-mat.str-el]}
  \BibitemShut {NoStop}%
\bibitem [{\citenamefont {{Campbell}}\ \emph {et~al.}(2019)\citenamefont
  {{Campbell}}, \citenamefont {{Brubaker}}, \citenamefont {{Roncaioli}},
  \citenamefont {{Saraf}}, \citenamefont {{Xiao}}, \citenamefont {{Chow}},
  \citenamefont {{Kenney-Benson}}, \citenamefont {{Popov}}, \citenamefont
  {{Zieve}}, \citenamefont {{Jeffries}},\ and\ \citenamefont
  {{Paglione}}}]{Campbell2019arxiv}%
  \BibitemOpen
  \bibfield  {author} {\bibinfo {author} {\bibfnamefont {D.~J.}\ \bibnamefont
  {{Campbell}}}, \bibinfo {author} {\bibfnamefont {Z.~E.}\ \bibnamefont
  {{Brubaker}}}, \bibinfo {author} {\bibfnamefont {C.}~\bibnamefont
  {{Roncaioli}}}, \bibinfo {author} {\bibfnamefont {P.}~\bibnamefont
  {{Saraf}}}, \bibinfo {author} {\bibfnamefont {Y.}~\bibnamefont {{Xiao}}},
  \bibinfo {author} {\bibfnamefont {P.}~\bibnamefont {{Chow}}}, \bibinfo
  {author} {\bibfnamefont {C.}~\bibnamefont {{Kenney-Benson}}}, \bibinfo
  {author} {\bibfnamefont {D.}~\bibnamefont {{Popov}}}, \bibinfo {author}
  {\bibfnamefont {R.~J.}\ \bibnamefont {{Zieve}}}, \bibinfo {author}
  {\bibfnamefont {J.~R.}\ \bibnamefont {{Jeffries}}}, \ and\ \bibinfo {author}
  {\bibfnamefont {J.}~\bibnamefont {{Paglione}}},\ }\href@noop {} {\bibfield
  {journal} {\bibinfo  {journal} {arXiv e-prints}\ ,\ \bibinfo {eid}
  {arXiv:1907.09017}} (\bibinfo {year} {2019})},\ \Eprint
  {http://arxiv.org/abs/1907.09017} {arXiv:1907.09017 [cond-mat.str-el]}
  \BibitemShut {NoStop}%
\bibitem [{\citenamefont {Kasaya}\ \emph
  {et~al.}(1991{\natexlab{a}})\citenamefont {Kasaya}, \citenamefont {Katoh},\
  and\ \citenamefont {Takegahara}}]{KASAYA1991797}%
  \BibitemOpen
  \bibfield  {author} {\bibinfo {author} {\bibfnamefont {M.}~\bibnamefont
  {Kasaya}}, \bibinfo {author} {\bibfnamefont {K.}~\bibnamefont {Katoh}}, \
  and\ \bibinfo {author} {\bibfnamefont {K.}~\bibnamefont {Takegahara}},\
  }\href {\doibase https://doi.org/10.1016/0038-1098(91)90623-4} {\bibfield
  {journal} {\bibinfo  {journal} {Solid State Communications}\ }\textbf
  {\bibinfo {volume} {78}},\ \bibinfo {pages} {797 } (\bibinfo {year}
  {1991}{\natexlab{a}})}\BibitemShut {NoStop}%
\bibitem [{\citenamefont {Katoh}\ and\ \citenamefont
  {Takabatake}(1998)}]{Katoh199822}%
  \BibitemOpen
  \bibfield  {author} {\bibinfo {author} {\bibfnamefont {K.}~\bibnamefont
  {Katoh}}\ and\ \bibinfo {author} {\bibfnamefont {T.}~\bibnamefont
  {Takabatake}},\ }\href {\doibase
  https://doi.org/10.1016/S0925-8388(97)00583-5} {\bibfield  {journal}
  {\bibinfo  {journal} {Journal of Alloys and Compounds}\ }\textbf {\bibinfo
  {volume} {268}},\ \bibinfo {pages} {22 } (\bibinfo {year}
  {1998})}\BibitemShut {NoStop}%
\bibitem [{\citenamefont {Dzero}\ \emph {et~al.}(2016)\citenamefont {Dzero},
  \citenamefont {Xia}, \citenamefont {Galitski},\ and\ \citenamefont
  {Coleman}}]{Dzero2016}%
  \BibitemOpen
  \bibfield  {author} {\bibinfo {author} {\bibfnamefont {M.}~\bibnamefont
  {Dzero}}, \bibinfo {author} {\bibfnamefont {J.}~\bibnamefont {Xia}}, \bibinfo
  {author} {\bibfnamefont {V.}~\bibnamefont {Galitski}}, \ and\ \bibinfo
  {author} {\bibfnamefont {P.}~\bibnamefont {Coleman}},\ }\href {\doibase
  10.1146/annurev-conmatphys-031214-014749} {\bibfield  {journal} {\bibinfo
  {journal} {Annual Review of Condensed Matter Physics}\ }\textbf {\bibinfo
  {volume} {7}},\ \bibinfo {pages} {249} (\bibinfo {year} {2016})}\BibitemShut
  {NoStop}%
\bibitem [{\citenamefont {Chang}\ \emph {et~al.}(2017)\citenamefont {Chang},
  \citenamefont {Erten},\ and\ \citenamefont {Coleman}}]{Chang2017}%
  \BibitemOpen
  \bibfield  {author} {\bibinfo {author} {\bibfnamefont {P.-Y.}\ \bibnamefont
  {Chang}}, \bibinfo {author} {\bibfnamefont {O.}~\bibnamefont {Erten}}, \ and\
  \bibinfo {author} {\bibfnamefont {P.}~\bibnamefont {Coleman}},\ }\href
  {http://dx.doi.org/10.1038/nphys4092} {\bibfield  {journal} {\bibinfo
  {journal} {Nature Physics}\ }\textbf {\bibinfo {volume} {13}},\ \bibinfo
  {pages} {794 EP } (\bibinfo {year} {2017})}\BibitemShut {NoStop}%
\bibitem [{\citenamefont {Jones}\ \emph {et~al.}(1998)\citenamefont {Jones},
  \citenamefont {Regan},\ and\ \citenamefont {DiSalvo}}]{PhysRevB.58.16057}%
  \BibitemOpen
  \bibfield  {author} {\bibinfo {author} {\bibfnamefont {C.~D.~W.}\
  \bibnamefont {Jones}}, \bibinfo {author} {\bibfnamefont {K.~A.}\ \bibnamefont
  {Regan}}, \ and\ \bibinfo {author} {\bibfnamefont {F.~J.}\ \bibnamefont
  {DiSalvo}},\ }\href {\doibase 10.1103/PhysRevB.58.16057} {\bibfield
  {journal} {\bibinfo  {journal} {Phys. Rev. B}\ }\textbf {\bibinfo {volume}
  {58}},\ \bibinfo {pages} {16057} (\bibinfo {year} {1998})}\BibitemShut
  {NoStop}%
\bibitem [{\citenamefont {Kasaya}\ \emph {et~al.}(1994)\citenamefont {Kasaya},
  \citenamefont {Katoh}, \citenamefont {Kohgi}, \citenamefont {Osakabe},\ and\
  \citenamefont {Sato}}]{KASAYA1994534}%
  \BibitemOpen
  \bibfield  {author} {\bibinfo {author} {\bibfnamefont {M.}~\bibnamefont
  {Kasaya}}, \bibinfo {author} {\bibfnamefont {K.}~\bibnamefont {Katoh}},
  \bibinfo {author} {\bibfnamefont {M.}~\bibnamefont {Kohgi}}, \bibinfo
  {author} {\bibfnamefont {T.}~\bibnamefont {Osakabe}}, \ and\ \bibinfo
  {author} {\bibfnamefont {N.}~\bibnamefont {Sato}},\ }\href {\doibase
  https://doi.org/10.1016/0921-4526(94)91896-1} {\bibfield  {journal} {\bibinfo
   {journal} {Physica B: Condensed Matter}\ }\textbf {\bibinfo {volume}
  {199-200}},\ \bibinfo {pages} {534 } (\bibinfo {year} {1994})}\BibitemShut
  {NoStop}%
\bibitem [{\citenamefont {Dwight}(1977)}]{Dwight:a14686}%
  \BibitemOpen
  \bibfield  {author} {\bibinfo {author} {\bibfnamefont {A.~E.}\ \bibnamefont
  {Dwight}},\ }\href {\doibase 10.1107/S0567740877006530} {\bibfield  {journal}
  {\bibinfo  {journal} {Acta Crystallographica Section B}\ }\textbf {\bibinfo
  {volume} {33}},\ \bibinfo {pages} {1579} (\bibinfo {year}
  {1977})}\BibitemShut {NoStop}%
\bibitem [{\citenamefont {Jones}\ \emph {et~al.}(1999)\citenamefont {Jones},
  \citenamefont {Regan},\ and\ \citenamefont {DiSalvo}}]{PhysRevB.60.5282}%
  \BibitemOpen
  \bibfield  {author} {\bibinfo {author} {\bibfnamefont {C.~D.~W.}\
  \bibnamefont {Jones}}, \bibinfo {author} {\bibfnamefont {K.~A.}\ \bibnamefont
  {Regan}}, \ and\ \bibinfo {author} {\bibfnamefont {F.~J.}\ \bibnamefont
  {DiSalvo}},\ }\href {\doibase 10.1103/PhysRevB.60.5282} {\bibfield  {journal}
  {\bibinfo  {journal} {Phys. Rev. B}\ }\textbf {\bibinfo {volume} {60}},\
  \bibinfo {pages} {5282} (\bibinfo {year} {1999})}\BibitemShut {NoStop}%
\bibitem [{\citenamefont {Takegahara}\ \emph {et~al.}(1993)\citenamefont
  {Takegahara}, \citenamefont {Harima}, \citenamefont {Kaneta},\ and\
  \citenamefont {Yanase}}]{doi:10.1143/JPSJ.62.2103}%
  \BibitemOpen
  \bibfield  {author} {\bibinfo {author} {\bibfnamefont {K.}~\bibnamefont
  {Takegahara}}, \bibinfo {author} {\bibfnamefont {H.}~\bibnamefont {Harima}},
  \bibinfo {author} {\bibfnamefont {Y.}~\bibnamefont {Kaneta}}, \ and\ \bibinfo
  {author} {\bibfnamefont {A.}~\bibnamefont {Yanase}},\ }\href {\doibase
  10.1143/JPSJ.62.2103} {\bibfield  {journal} {\bibinfo  {journal} {Journal of
  the Physical Society of Japan}\ }\textbf {\bibinfo {volume} {62}},\ \bibinfo
  {pages} {2103} (\bibinfo {year} {1993})}\BibitemShut {NoStop}%
\bibitem [{\citenamefont {Blaha}\ \emph {et~al.}(2001)\citenamefont {Blaha},
  \citenamefont {Schwarz}, \citenamefont {Madsen}, \citenamefont {Kvasnicka},\
  and\ \citenamefont {Luitz}}]{wien2k}%
  \BibitemOpen
  \bibfield  {author} {\bibinfo {author} {\bibfnamefont {P.}~\bibnamefont
  {Blaha}}, \bibinfo {author} {\bibfnamefont {K.}~\bibnamefont {Schwarz}},
  \bibinfo {author} {\bibfnamefont {G.-K.-H.}\ \bibnamefont {Madsen}}, \bibinfo
  {author} {\bibfnamefont {D.}~\bibnamefont {Kvasnicka}}, \ and\ \bibinfo
  {author} {\bibfnamefont {J.}~\bibnamefont {Luitz}},\ }\href@noop {}
  {\bibfield  {journal} {\bibinfo  {journal} {Vienna University of Technology,
  Austria}\ } (\bibinfo {year} {2001})},\ \bibinfo {note} {iSBN
  3-9501031-1-2}\BibitemShut {NoStop}%
\bibitem [{\citenamefont {Marzari}\ \emph {et~al.}(2012)\citenamefont
  {Marzari}, \citenamefont {Mostofi}, \citenamefont {Yates}, \citenamefont
  {Souza},\ and\ \citenamefont {Vanderbilt}}]{RevModPhys.84.1419}%
  \BibitemOpen
  \bibfield  {author} {\bibinfo {author} {\bibfnamefont {N.}~\bibnamefont
  {Marzari}}, \bibinfo {author} {\bibfnamefont {A.~A.}\ \bibnamefont
  {Mostofi}}, \bibinfo {author} {\bibfnamefont {J.~R.}\ \bibnamefont {Yates}},
  \bibinfo {author} {\bibfnamefont {I.}~\bibnamefont {Souza}}, \ and\ \bibinfo
  {author} {\bibfnamefont {D.}~\bibnamefont {Vanderbilt}},\ }\href {\doibase
  10.1103/RevModPhys.84.1419} {\bibfield  {journal} {\bibinfo  {journal} {Rev.
  Mod. Phys.}\ }\textbf {\bibinfo {volume} {84}},\ \bibinfo {pages} {1419}
  (\bibinfo {year} {2012})}\BibitemShut {NoStop}%
\bibitem [{\citenamefont {Mostofi}\ \emph {et~al.}(2008)\citenamefont
  {Mostofi}, \citenamefont {Yates}, \citenamefont {Lee}, \citenamefont {Souza},
  \citenamefont {Vanderbilt},\ and\ \citenamefont {Marzari}}]{wannier90}%
  \BibitemOpen
  \bibfield  {author} {\bibinfo {author} {\bibfnamefont {A.~A.}\ \bibnamefont
  {Mostofi}}, \bibinfo {author} {\bibfnamefont {J.~R.}\ \bibnamefont {Yates}},
  \bibinfo {author} {\bibfnamefont {Y.-S.}\ \bibnamefont {Lee}}, \bibinfo
  {author} {\bibfnamefont {I.}~\bibnamefont {Souza}}, \bibinfo {author}
  {\bibfnamefont {D.}~\bibnamefont {Vanderbilt}}, \ and\ \bibinfo {author}
  {\bibfnamefont {N.}~\bibnamefont {Marzari}},\ }\href {\doibase
  https://doi.org/10.1016/j.cpc.2007.11.016} {\bibfield  {journal} {\bibinfo
  {journal} {Computer Physics Communications}\ }\textbf {\bibinfo {volume}
  {178}},\ \bibinfo {pages} {685 } (\bibinfo {year} {2008})}\BibitemShut
  {NoStop}%
\bibitem [{\citenamefont {Kunes}\ \emph {et~al.}(2010)\citenamefont {Kunes},
  \citenamefont {Arita}, \citenamefont {Wissgott}, \citenamefont {Toschi},
  \citenamefont {Ikeda},\ and\ \citenamefont {Held}}]{wien2wannier}%
  \BibitemOpen
  \bibfield  {author} {\bibinfo {author} {\bibfnamefont {J.}~\bibnamefont
  {Kunes}}, \bibinfo {author} {\bibfnamefont {R.}~\bibnamefont {Arita}},
  \bibinfo {author} {\bibfnamefont {P.}~\bibnamefont {Wissgott}}, \bibinfo
  {author} {\bibfnamefont {A.}~\bibnamefont {Toschi}}, \bibinfo {author}
  {\bibfnamefont {H.}~\bibnamefont {Ikeda}}, \ and\ \bibinfo {author}
  {\bibfnamefont {K.}~\bibnamefont {Held}},\ }\href {\doibase
  http://dx.doi.org/10.1016/j.cpc.2010.08.005} {\bibfield  {journal} {\bibinfo
  {journal} {Computer Physics Communications}\ }\textbf {\bibinfo {volume}
  {181}},\ \bibinfo {pages} {1888 } (\bibinfo {year} {2010})}\BibitemShut
  {NoStop}%
\bibitem [{\citenamefont {Fernandez~Afonso}(2019)}]{JuanPhD}%
  \BibitemOpen
  \bibfield  {author} {\bibinfo {author} {\bibfnamefont {J.}~\bibnamefont
  {Fernandez~Afonso}},\ }\href@noop {} {\emph {\bibinfo {title} {Excitonic
  condensation in strongly correlated materials}}}\ (\bibinfo  {publisher} {PhD
  thesis, TU Wien},\ \bibinfo {year} {2019})\BibitemShut {NoStop}%
\bibitem [{\citenamefont {Haule}\ \emph {et~al.}(2010)\citenamefont {Haule},
  \citenamefont {Yee},\ and\ \citenamefont {Kim}}]{PhysRevB.81.195107}%
  \BibitemOpen
  \bibfield  {author} {\bibinfo {author} {\bibfnamefont {K.}~\bibnamefont
  {Haule}}, \bibinfo {author} {\bibfnamefont {C.-H.}\ \bibnamefont {Yee}}, \
  and\ \bibinfo {author} {\bibfnamefont {K.}~\bibnamefont {Kim}},\ }\href
  {\doibase 10.1103/PhysRevB.81.195107} {\bibfield  {journal} {\bibinfo
  {journal} {Phys. Rev. B}\ }\textbf {\bibinfo {volume} {81}},\ \bibinfo
  {pages} {195107} (\bibinfo {year} {2010})}\BibitemShut {NoStop}%
\bibitem [{\citenamefont {Held}\ \emph {et~al.}(2001)\citenamefont {Held},
  \citenamefont {McMahan},\ and\ \citenamefont
  {Scalettar}}]{PhysRevLett.87.276404}%
  \BibitemOpen
  \bibfield  {author} {\bibinfo {author} {\bibfnamefont {K.}~\bibnamefont
  {Held}}, \bibinfo {author} {\bibfnamefont {A.~K.}\ \bibnamefont {McMahan}}, \
  and\ \bibinfo {author} {\bibfnamefont {R.~T.}\ \bibnamefont {Scalettar}},\
  }\href {\doibase 10.1103/PhysRevLett.87.276404} {\bibfield  {journal}
  {\bibinfo  {journal} {Phys. Rev. Lett.}\ }\textbf {\bibinfo {volume} {87}},\
  \bibinfo {pages} {276404} (\bibinfo {year} {2001})}\BibitemShut {NoStop}%
\bibitem [{\citenamefont {Amadon}\ \emph {et~al.}(2006)\citenamefont {Amadon},
  \citenamefont {Biermann}, \citenamefont {Georges},\ and\ \citenamefont
  {Aryasetiawan}}]{amadon:066402}%
  \BibitemOpen
  \bibfield  {author} {\bibinfo {author} {\bibfnamefont {B.}~\bibnamefont
  {Amadon}}, \bibinfo {author} {\bibfnamefont {S.}~\bibnamefont {Biermann}},
  \bibinfo {author} {\bibfnamefont {A.}~\bibnamefont {Georges}}, \ and\
  \bibinfo {author} {\bibfnamefont {F.}~\bibnamefont {Aryasetiawan}},\ }\href
  {\doibase 10.1103/PhysRevLett.96.066402} {\bibfield  {journal} {\bibinfo
  {journal} {Phys. Rev. Lett.}\ }\textbf {\bibinfo {volume} {96}},\ \bibinfo
  {eid} {066402} (\bibinfo {year} {2006})}\BibitemShut {NoStop}%
\bibitem [{\citenamefont {Tomczak}\ \emph {et~al.}(2013)\citenamefont
  {Tomczak}, \citenamefont {Pourovskii}, \citenamefont {Vaugier}, \citenamefont
  {Georges},\ and\ \citenamefont {Biermann}}]{jmt_cesf}%
  \BibitemOpen
  \bibfield  {author} {\bibinfo {author} {\bibfnamefont {J.~M.}\ \bibnamefont
  {Tomczak}}, \bibinfo {author} {\bibfnamefont {L.~V.}\ \bibnamefont
  {Pourovskii}}, \bibinfo {author} {\bibfnamefont {L.}~\bibnamefont {Vaugier}},
  \bibinfo {author} {\bibfnamefont {A.}~\bibnamefont {Georges}}, \ and\
  \bibinfo {author} {\bibfnamefont {S.}~\bibnamefont {Biermann}},\ }\href
  {\doibase 10.1073/pnas.1215066110} {\bibfield  {journal} {\bibinfo  {journal}
  {Proc. Natl. Acad. Sci. USA}\ }\textbf {\bibinfo {volume} {110}},\ \bibinfo
  {pages} {904} (\bibinfo {year} {2013})}\BibitemShut {NoStop}%
\bibitem [{\citenamefont {Pourovskii}\ \emph {et~al.}(2014)\citenamefont
  {Pourovskii}, \citenamefont {Hansmann}, \citenamefont {Ferrero},\ and\
  \citenamefont {Georges}}]{PhysRevLett.112.106407}%
  \BibitemOpen
  \bibfield  {author} {\bibinfo {author} {\bibfnamefont {L.~V.}\ \bibnamefont
  {Pourovskii}}, \bibinfo {author} {\bibfnamefont {P.}~\bibnamefont
  {Hansmann}}, \bibinfo {author} {\bibfnamefont {M.}~\bibnamefont {Ferrero}}, \
  and\ \bibinfo {author} {\bibfnamefont {A.}~\bibnamefont {Georges}},\ }\href
  {\doibase 10.1103/PhysRevLett.112.106407} {\bibfield  {journal} {\bibinfo
  {journal} {Phys. Rev. Lett.}\ }\textbf {\bibinfo {volume} {112}},\ \bibinfo
  {pages} {106407} (\bibinfo {year} {2014})}\BibitemShut {NoStop}%
\bibitem [{\citenamefont {Shick}\ \emph {et~al.}(2015)\citenamefont {Shick},
  \citenamefont {Havela}, \citenamefont {Lichtenstein},\ and\ \citenamefont
  {Katsnelson}}]{Shick2015}%
  \BibitemOpen
  \bibfield  {author} {\bibinfo {author} {\bibfnamefont {A.~B.}\ \bibnamefont
  {Shick}}, \bibinfo {author} {\bibfnamefont {L.}~\bibnamefont {Havela}},
  \bibinfo {author} {\bibfnamefont {A.~I.}\ \bibnamefont {Lichtenstein}}, \
  and\ \bibinfo {author} {\bibfnamefont {M.~I.}\ \bibnamefont {Katsnelson}},\
  }\href {http://dx.doi.org/10.1038/srep15429} {\bibfield  {journal} {\bibinfo
  {journal} {Scientific Reports}\ }\textbf {\bibinfo {volume} {5}},\ \bibinfo
  {pages} {15429 EP } (\bibinfo {year} {2015})}\BibitemShut {NoStop}%
\bibitem [{\citenamefont {Wissgott}\ and\ \citenamefont
  {Held}(2016)}]{Wissgott2016}%
  \BibitemOpen
  \bibfield  {author} {\bibinfo {author} {\bibfnamefont {P.}~\bibnamefont
  {Wissgott}}\ and\ \bibinfo {author} {\bibfnamefont {K.}~\bibnamefont
  {Held}},\ }\href {\doibase 10.1140/epjb/e2015-60753-5} {\bibfield  {journal}
  {\bibinfo  {journal} {Eur. Phys. J. B}\ }\textbf {\bibinfo {volume} {89}},\
  \bibinfo {pages} {5} (\bibinfo {year} {2016})}\BibitemShut {NoStop}%
\bibitem [{\citenamefont {Plekhanov}\ \emph {et~al.}(2018)\citenamefont
  {Plekhanov}, \citenamefont {Hasnip}, \citenamefont {Sacksteder},
  \citenamefont {Probert}, \citenamefont {Clark}, \citenamefont {Refson},\ and\
  \citenamefont {Weber}}]{PhysRevB.98.075129}%
  \BibitemOpen
  \bibfield  {author} {\bibinfo {author} {\bibfnamefont {E.}~\bibnamefont
  {Plekhanov}}, \bibinfo {author} {\bibfnamefont {P.}~\bibnamefont {Hasnip}},
  \bibinfo {author} {\bibfnamefont {V.}~\bibnamefont {Sacksteder}}, \bibinfo
  {author} {\bibfnamefont {M.}~\bibnamefont {Probert}}, \bibinfo {author}
  {\bibfnamefont {S.~J.}\ \bibnamefont {Clark}}, \bibinfo {author}
  {\bibfnamefont {K.}~\bibnamefont {Refson}}, \ and\ \bibinfo {author}
  {\bibfnamefont {C.}~\bibnamefont {Weber}},\ }\href {\doibase
  10.1103/PhysRevB.98.075129} {\bibfield  {journal} {\bibinfo  {journal} {Phys.
  Rev. B}\ }\textbf {\bibinfo {volume} {98}},\ \bibinfo {pages} {075129}
  (\bibinfo {year} {2018})}\BibitemShut {NoStop}%
\bibitem [{\citenamefont {Seong}\ \emph {et~al.}(2019)\citenamefont {Seong},
  \citenamefont {Kim}, \citenamefont {Lee}, \citenamefont {Kang}, \citenamefont
  {Nam}, \citenamefont {Min}, \citenamefont {Yoshino}, \citenamefont
  {Takabatake}, \citenamefont {Denlinger},\ and\ \citenamefont
  {Kang}}]{PhysRevB.100.035121}%
  \BibitemOpen
  \bibfield  {author} {\bibinfo {author} {\bibfnamefont {S.}~\bibnamefont
  {Seong}}, \bibinfo {author} {\bibfnamefont {K.}~\bibnamefont {Kim}}, \bibinfo
  {author} {\bibfnamefont {E.}~\bibnamefont {Lee}}, \bibinfo {author}
  {\bibfnamefont {C.-J.}\ \bibnamefont {Kang}}, \bibinfo {author}
  {\bibfnamefont {T.}~\bibnamefont {Nam}}, \bibinfo {author} {\bibfnamefont
  {B.~I.}\ \bibnamefont {Min}}, \bibinfo {author} {\bibfnamefont
  {T.}~\bibnamefont {Yoshino}}, \bibinfo {author} {\bibfnamefont
  {T.}~\bibnamefont {Takabatake}}, \bibinfo {author} {\bibfnamefont {J.~D.}\
  \bibnamefont {Denlinger}}, \ and\ \bibinfo {author} {\bibfnamefont {J.-S.}\
  \bibnamefont {Kang}},\ }\href {\doibase 10.1103/PhysRevB.100.035121}
  {\bibfield  {journal} {\bibinfo  {journal} {Phys. Rev. B}\ }\textbf {\bibinfo
  {volume} {100}},\ \bibinfo {pages} {035121} (\bibinfo {year}
  {2019})}\BibitemShut {NoStop}%
\bibitem [{\citenamefont {Zwicknagl}(2016)}]{0034-4885-79-12-124501}%
  \BibitemOpen
  \bibfield  {author} {\bibinfo {author} {\bibfnamefont {G.}~\bibnamefont
  {Zwicknagl}},\ }\href {http://stacks.iop.org/0034-4885/79/i=12/a=124501}
  {\bibfield  {journal} {\bibinfo  {journal} {Rep. Prog. Phys.}\ }\textbf
  {\bibinfo {volume} {79}},\ \bibinfo {pages} {124501} (\bibinfo {year}
  {2016})}\BibitemShut {NoStop}%
\bibitem [{\citenamefont {Hafiz}\ \emph {et~al.}(2018)\citenamefont {Hafiz},
  \citenamefont {Khair}, \citenamefont {Choi}, \citenamefont {Mueen},
  \citenamefont {Bansil}, \citenamefont {Eidenbenz}, \citenamefont {Wills},
  \citenamefont {Zhu}, \citenamefont {Balatsky},\ and\ \citenamefont
  {Ahmed}}]{Hafiz2018}%
  \BibitemOpen
  \bibfield  {author} {\bibinfo {author} {\bibfnamefont {H.}~\bibnamefont
  {Hafiz}}, \bibinfo {author} {\bibfnamefont {A.~I.}\ \bibnamefont {Khair}},
  \bibinfo {author} {\bibfnamefont {H.}~\bibnamefont {Choi}}, \bibinfo {author}
  {\bibfnamefont {A.}~\bibnamefont {Mueen}}, \bibinfo {author} {\bibfnamefont
  {A.}~\bibnamefont {Bansil}}, \bibinfo {author} {\bibfnamefont
  {S.}~\bibnamefont {Eidenbenz}}, \bibinfo {author} {\bibfnamefont
  {J.}~\bibnamefont {Wills}}, \bibinfo {author} {\bibfnamefont {J.-X.}\
  \bibnamefont {Zhu}}, \bibinfo {author} {\bibfnamefont {A.~V.}\ \bibnamefont
  {Balatsky}}, \ and\ \bibinfo {author} {\bibfnamefont {T.}~\bibnamefont
  {Ahmed}},\ }\href {\doibase 10.1038/s41524-018-0120-9} {\bibfield  {journal}
  {\bibinfo  {journal} {npj Computational Materials}\ }\textbf {\bibinfo
  {volume} {4}},\ \bibinfo {pages} {63} (\bibinfo {year} {2018})}\BibitemShut
  {NoStop}%
\bibitem [{Note1()}]{Note1}%
  \BibitemOpen
  \bibinfo {note} {Actually, there exists an exact mapping of the PAM to the
  KLM beyond the Kondo regime\cite {PhysRevB.65.212303}.}\BibitemShut {Stop}%
\bibitem [{\citenamefont {Fuggle}\ \emph {et~al.}(1983)\citenamefont {Fuggle},
  \citenamefont {Hillebrecht}, \citenamefont {Zo\l{}nierek}, \citenamefont
  {L\"asser}, \citenamefont {Freiburg}, \citenamefont {Gunnarsson},\ and\
  \citenamefont {Sch\"onhammer}}]{PhysRevB.27.7330}%
  \BibitemOpen
  \bibfield  {author} {\bibinfo {author} {\bibfnamefont {J.~C.}\ \bibnamefont
  {Fuggle}}, \bibinfo {author} {\bibfnamefont {F.~U.}\ \bibnamefont
  {Hillebrecht}}, \bibinfo {author} {\bibfnamefont {Z.}~\bibnamefont
  {Zo\l{}nierek}}, \bibinfo {author} {\bibfnamefont {R.}~\bibnamefont
  {L\"asser}}, \bibinfo {author} {\bibfnamefont {C.}~\bibnamefont {Freiburg}},
  \bibinfo {author} {\bibfnamefont {O.}~\bibnamefont {Gunnarsson}}, \ and\
  \bibinfo {author} {\bibfnamefont {K.}~\bibnamefont {Sch\"onhammer}},\ }\href
  {\doibase 10.1103/PhysRevB.27.7330} {\bibfield  {journal} {\bibinfo
  {journal} {Phys. Rev. B}\ }\textbf {\bibinfo {volume} {27}},\ \bibinfo
  {pages} {7330} (\bibinfo {year} {1983})}\BibitemShut {NoStop}%
\bibitem [{\citenamefont {Strigari}\ \emph {et~al.}(2012)\citenamefont
  {Strigari}, \citenamefont {Willers}, \citenamefont {Muro}, \citenamefont
  {Yutani}, \citenamefont {Takabatake}, \citenamefont {Hu}, \citenamefont
  {Chin}, \citenamefont {Agrestini}, \citenamefont {Lin}, \citenamefont {Chen},
  \citenamefont {Tanaka}, \citenamefont {Haverkort}, \citenamefont {Tjeng},\
  and\ \citenamefont {Severing}}]{PhysRevB.86.081105}%
  \BibitemOpen
  \bibfield  {author} {\bibinfo {author} {\bibfnamefont {F.}~\bibnamefont
  {Strigari}}, \bibinfo {author} {\bibfnamefont {T.}~\bibnamefont {Willers}},
  \bibinfo {author} {\bibfnamefont {Y.}~\bibnamefont {Muro}}, \bibinfo {author}
  {\bibfnamefont {K.}~\bibnamefont {Yutani}}, \bibinfo {author} {\bibfnamefont
  {T.}~\bibnamefont {Takabatake}}, \bibinfo {author} {\bibfnamefont
  {Z.}~\bibnamefont {Hu}}, \bibinfo {author} {\bibfnamefont {Y.-Y.}\
  \bibnamefont {Chin}}, \bibinfo {author} {\bibfnamefont {S.}~\bibnamefont
  {Agrestini}}, \bibinfo {author} {\bibfnamefont {H.-J.}\ \bibnamefont {Lin}},
  \bibinfo {author} {\bibfnamefont {C.~T.}\ \bibnamefont {Chen}}, \bibinfo
  {author} {\bibfnamefont {A.}~\bibnamefont {Tanaka}}, \bibinfo {author}
  {\bibfnamefont {M.~W.}\ \bibnamefont {Haverkort}}, \bibinfo {author}
  {\bibfnamefont {L.~H.}\ \bibnamefont {Tjeng}}, \ and\ \bibinfo {author}
  {\bibfnamefont {A.}~\bibnamefont {Severing}},\ }\href {\doibase
  10.1103/PhysRevB.86.081105} {\bibfield  {journal} {\bibinfo  {journal} {Phys.
  Rev. B}\ }\textbf {\bibinfo {volume} {86}},\ \bibinfo {pages} {081105}
  (\bibinfo {year} {2012})}\BibitemShut {NoStop}%
\bibitem [{\citenamefont {Strigari}\ \emph {et~al.}(2015)\citenamefont
  {Strigari}, \citenamefont {Sundermann}, \citenamefont {Muro}, \citenamefont
  {Yutani}, \citenamefont {Takabatake}, \citenamefont {Tsuei}, \citenamefont
  {Liao}, \citenamefont {Tanaka}, \citenamefont {Thalmeier}, \citenamefont
  {Haverkort}, \citenamefont {Tjeng},\ and\ \citenamefont
  {Severing}}]{STRIGARI201556}%
  \BibitemOpen
  \bibfield  {author} {\bibinfo {author} {\bibfnamefont {F.}~\bibnamefont
  {Strigari}}, \bibinfo {author} {\bibfnamefont {M.}~\bibnamefont
  {Sundermann}}, \bibinfo {author} {\bibfnamefont {Y.}~\bibnamefont {Muro}},
  \bibinfo {author} {\bibfnamefont {K.}~\bibnamefont {Yutani}}, \bibinfo
  {author} {\bibfnamefont {T.}~\bibnamefont {Takabatake}}, \bibinfo {author}
  {\bibfnamefont {K.-D.}\ \bibnamefont {Tsuei}}, \bibinfo {author}
  {\bibfnamefont {Y.}~\bibnamefont {Liao}}, \bibinfo {author} {\bibfnamefont
  {A.}~\bibnamefont {Tanaka}}, \bibinfo {author} {\bibfnamefont
  {P.}~\bibnamefont {Thalmeier}}, \bibinfo {author} {\bibfnamefont
  {M.}~\bibnamefont {Haverkort}}, \bibinfo {author} {\bibfnamefont
  {L.}~\bibnamefont {Tjeng}}, \ and\ \bibinfo {author} {\bibfnamefont
  {A.}~\bibnamefont {Severing}},\ }\href {\doibase
  https://doi.org/10.1016/j.elspec.2015.01.004} {\bibfield  {journal} {\bibinfo
   {journal} {Journal of Electron Spectroscopy and Related Phenomena}\ }\textbf
  {\bibinfo {volume} {199}},\ \bibinfo {pages} {56 } (\bibinfo {year}
  {2015})}\BibitemShut {NoStop}%
\bibitem [{Note2()}]{Note2}%
  \BibitemOpen
  \bibinfo {note} {See, however, e.g., Refs.~\protect \rev@citealpnum
  {doi:10.1080/01411590701228703,PhysRevLett.112.106407} for multiplet effects
  in Ce-compounds under external pressure.}\BibitemShut {Stop}%
\bibitem [{Note3()}]{Note3}%
  \BibitemOpen
  \bibinfo {note} {All relative changes are slightly larger when using LDA
  instead of PBE.}\BibitemShut {Stop}%
\bibitem [{\citenamefont {Takabatake}\ \emph {et~al.}(1993)\citenamefont
  {Takabatake}, \citenamefont {Iwasaki}, \citenamefont {Nakamoto},
  \citenamefont {Fujii}, \citenamefont {Nakotte}, \citenamefont {de~Boer},\
  and\ \citenamefont {Sechovský}}]{TAKABATAKE1993108}%
  \BibitemOpen
  \bibfield  {author} {\bibinfo {author} {\bibfnamefont {T.}~\bibnamefont
  {Takabatake}}, \bibinfo {author} {\bibfnamefont {H.}~\bibnamefont {Iwasaki}},
  \bibinfo {author} {\bibfnamefont {G.}~\bibnamefont {Nakamoto}}, \bibinfo
  {author} {\bibfnamefont {H.}~\bibnamefont {Fujii}}, \bibinfo {author}
  {\bibfnamefont {H.}~\bibnamefont {Nakotte}}, \bibinfo {author} {\bibfnamefont
  {F.}~\bibnamefont {de~Boer}}, \ and\ \bibinfo {author} {\bibfnamefont
  {V.}~\bibnamefont {Sechovský}},\ }\href {\doibase
  https://doi.org/10.1016/0921-4526(93)90061-A} {\bibfield  {journal} {\bibinfo
   {journal} {Physica B: Condensed Matter}\ }\textbf {\bibinfo {volume}
  {183}},\ \bibinfo {pages} {108 } (\bibinfo {year} {1993})}\BibitemShut
  {NoStop}%
\bibitem [{\citenamefont {Malik}\ \emph {et~al.}(1989)\citenamefont {Malik},
  \citenamefont {Adroja}, \citenamefont {Dhar}, \citenamefont
  {Vijayaraghavan},\ and\ \citenamefont {Padalia}}]{PhysRevB.40.2414}%
  \BibitemOpen
  \bibfield  {author} {\bibinfo {author} {\bibfnamefont {S.~K.}\ \bibnamefont
  {Malik}}, \bibinfo {author} {\bibfnamefont {D.~T.}\ \bibnamefont {Adroja}},
  \bibinfo {author} {\bibfnamefont {S.~K.}\ \bibnamefont {Dhar}}, \bibinfo
  {author} {\bibfnamefont {R.}~\bibnamefont {Vijayaraghavan}}, \ and\ \bibinfo
  {author} {\bibfnamefont {B.~D.}\ \bibnamefont {Padalia}},\ }\href {\doibase
  10.1103/PhysRevB.40.2414} {\bibfield  {journal} {\bibinfo  {journal} {Phys.
  Rev. B}\ }\textbf {\bibinfo {volume} {40}},\ \bibinfo {pages} {2414}
  (\bibinfo {year} {1989})}\BibitemShut {NoStop}%
\bibitem [{\citenamefont {Breuer}\ \emph {et~al.}(1998)\citenamefont {Breuer},
  \citenamefont {Messerli}, \citenamefont {Purdie}, \citenamefont {Garnier},
  \citenamefont {Hengsberger}, \citenamefont {Panaccione}, \citenamefont
  {Baer}, \citenamefont {Takahashi}, \citenamefont {Yoshii}, \citenamefont
  {Kasaya}, \citenamefont {Katoh},\ and\ \citenamefont
  {Takabatake}}]{0295-5075-41-5-565}%
  \BibitemOpen
  \bibfield  {author} {\bibinfo {author} {\bibfnamefont {K.}~\bibnamefont
  {Breuer}}, \bibinfo {author} {\bibfnamefont {S.}~\bibnamefont {Messerli}},
  \bibinfo {author} {\bibfnamefont {D.}~\bibnamefont {Purdie}}, \bibinfo
  {author} {\bibfnamefont {M.}~\bibnamefont {Garnier}}, \bibinfo {author}
  {\bibfnamefont {M.}~\bibnamefont {Hengsberger}}, \bibinfo {author}
  {\bibfnamefont {G.}~\bibnamefont {Panaccione}}, \bibinfo {author}
  {\bibfnamefont {Y.}~\bibnamefont {Baer}}, \bibinfo {author} {\bibfnamefont
  {T.}~\bibnamefont {Takahashi}}, \bibinfo {author} {\bibfnamefont
  {S.}~\bibnamefont {Yoshii}}, \bibinfo {author} {\bibfnamefont
  {M.}~\bibnamefont {Kasaya}}, \bibinfo {author} {\bibfnamefont
  {K.}~\bibnamefont {Katoh}}, \ and\ \bibinfo {author} {\bibfnamefont
  {T.}~\bibnamefont {Takabatake}},\ }\href
  {http://stacks.iop.org/0295-5075/41/i=5/a=565} {\bibfield  {journal}
  {\bibinfo  {journal} {EPL (Europhysics Letters)}\ }\textbf {\bibinfo {volume}
  {41}},\ \bibinfo {pages} {565} (\bibinfo {year} {1998})}\BibitemShut
  {NoStop}%
\bibitem [{\citenamefont {Takeda}\ \emph {et~al.}(1999)\citenamefont {Takeda},
  \citenamefont {Arita}, \citenamefont {Sato}, \citenamefont {Shimada},
  \citenamefont {Namatame}, \citenamefont {Taniguchi}, \citenamefont {Katoh},
  \citenamefont {Iga},\ and\ \citenamefont {Takabatake}}]{Takeda1999721}%
  \BibitemOpen
  \bibfield  {author} {\bibinfo {author} {\bibfnamefont {Y.}~\bibnamefont
  {Takeda}}, \bibinfo {author} {\bibfnamefont {M.}~\bibnamefont {Arita}},
  \bibinfo {author} {\bibfnamefont {H.}~\bibnamefont {Sato}}, \bibinfo {author}
  {\bibfnamefont {K.}~\bibnamefont {Shimada}}, \bibinfo {author} {\bibfnamefont
  {H.}~\bibnamefont {Namatame}}, \bibinfo {author} {\bibfnamefont
  {M.}~\bibnamefont {Taniguchi}}, \bibinfo {author} {\bibfnamefont
  {K.}~\bibnamefont {Katoh}}, \bibinfo {author} {\bibfnamefont
  {F.}~\bibnamefont {Iga}}, \ and\ \bibinfo {author} {\bibfnamefont
  {T.}~\bibnamefont {Takabatake}},\ }\href {\doibase
  https://doi.org/10.1016/S0368-2048(98)00406-X} {\bibfield  {journal}
  {\bibinfo  {journal} {J Electron Spectrosc}\ }\textbf {\bibinfo {volume}
  {101–103}},\ \bibinfo {pages} {721 } (\bibinfo {year} {1999})}\BibitemShut
  {NoStop}%
\bibitem [{Note4()}]{Note4}%
  \BibitemOpen
  \bibinfo {note} {See also Ref.~\protect \rev@citealpnum {NGCS} where such a
  construction helped distinguishing the gap nature in covalent FeSi and ionic
  LaCoO$_3$.}\BibitemShut {Stop}%
\bibitem [{\citenamefont {Bondi}(1964)}]{Bondi1964}%
  \BibitemOpen
  \bibfield  {author} {\bibinfo {author} {\bibfnamefont {A.}~\bibnamefont
  {Bondi}},\ }\href {\doibase 10.1021/j100785a001} {\bibfield  {journal}
  {\bibinfo  {journal} {The Journal of Physical Chemistry}\ }\textbf {\bibinfo
  {volume} {68}},\ \bibinfo {pages} {441} (\bibinfo {year} {1964})},\ \Eprint
  {http://arxiv.org/abs/https://doi.org/10.1021/j100785a001}
  {https://doi.org/10.1021/j100785a001} \BibitemShut {NoStop}%
\bibitem [{Gre(1997)}]{Greenwood1997}%
  \BibitemOpen
  in\ \href {\doibase https://doi.org/10.1016/B978-0-7506-3365-9.50033-X}
  {\emph {\bibinfo {booktitle} {Chemistry of the Elements (Second Edition)}}},\
  \bibinfo {editor} {edited by\ \bibinfo {editor} {\bibfnamefont
  {N.}~\bibnamefont {Greenwood}}\ and\ \bibinfo {editor} {\bibfnamefont
  {A.}~\bibnamefont {Earnshaw}}}\ (\bibinfo  {publisher}
  {Butterworth-Heinemann},\ \bibinfo {address} {Oxford},\ \bibinfo {year}
  {1997})\ \bibinfo {edition} {second edition}\ ed.,\ pp.\ \bibinfo {pages}
  {1144 -- 1172}\BibitemShut {NoStop}%
\bibitem [{\citenamefont {Tomczak}\ \emph {et~al.}(2009)\citenamefont
  {Tomczak}, \citenamefont {Miyake}, \citenamefont {Sakuma},\ and\
  \citenamefont {Aryasetiawan}}]{jmt_wannier}%
  \BibitemOpen
  \bibfield  {author} {\bibinfo {author} {\bibfnamefont {J.~M.}\ \bibnamefont
  {Tomczak}}, \bibinfo {author} {\bibfnamefont {T.}~\bibnamefont {Miyake}},
  \bibinfo {author} {\bibfnamefont {R.}~\bibnamefont {Sakuma}}, \ and\ \bibinfo
  {author} {\bibfnamefont {F.}~\bibnamefont {Aryasetiawan}},\ }\href {\doibase
  10.1103/PhysRevB.79.235133} {\bibfield  {journal} {\bibinfo  {journal} {Phys.
  Rev. B}\ }\textbf {\bibinfo {volume} {79}},\ \bibinfo {eid} {235133}
  (\bibinfo {year} {2009})},\ \bibinfo {note} {preprint
  arXiv:0906.4398}\BibitemShut {NoStop}%
\bibitem [{\citenamefont {Tomczak}\ \emph {et~al.}(2010)\citenamefont
  {Tomczak}, \citenamefont {Miyake},\ and\ \citenamefont
  {Aryasetiawan}}]{jmt_mno}%
  \BibitemOpen
  \bibfield  {author} {\bibinfo {author} {\bibfnamefont {J.~M.}\ \bibnamefont
  {Tomczak}}, \bibinfo {author} {\bibfnamefont {T.}~\bibnamefont {Miyake}}, \
  and\ \bibinfo {author} {\bibfnamefont {F.}~\bibnamefont {Aryasetiawan}},\
  }\href {\doibase 10.1103/PhysRevB.81.115116} {\bibfield  {journal} {\bibinfo
  {journal} {Phys. Rev. B}\ }\textbf {\bibinfo {volume} {81}},\ \bibinfo
  {pages} {115116} (\bibinfo {year} {2010})},\ \bibinfo {note} {preprint
  arXiv:1006.0565}\BibitemShut {NoStop}%
\bibitem [{\citenamefont {Kasaya}\ \emph {et~al.}(1988)\citenamefont {Kasaya},
  \citenamefont {Tani}, \citenamefont {Iga},\ and\ \citenamefont
  {Kasuya}}]{KASAYA1988278}%
  \BibitemOpen
  \bibfield  {author} {\bibinfo {author} {\bibfnamefont {M.}~\bibnamefont
  {Kasaya}}, \bibinfo {author} {\bibfnamefont {T.}~\bibnamefont {Tani}},
  \bibinfo {author} {\bibfnamefont {F.}~\bibnamefont {Iga}}, \ and\ \bibinfo
  {author} {\bibfnamefont {T.}~\bibnamefont {Kasuya}},\ }\href {\doibase
  https://doi.org/10.1016/0304-8853(88)90395-2} {\bibfield  {journal} {\bibinfo
   {journal} {Journal of Magnetism and Magnetic Materials}\ }\textbf {\bibinfo
  {volume} {76-77}},\ \bibinfo {pages} {278 } (\bibinfo {year}
  {1988})}\BibitemShut {NoStop}%
\bibitem [{\citenamefont {Iga}\ \emph {et~al.}(1993)\citenamefont {Iga},
  \citenamefont {Kasaya}, \citenamefont {Suzuki}, \citenamefont {Okayama},
  \citenamefont {Takahashi},\ and\ \citenamefont {Mori}}]{IGA1993419}%
  \BibitemOpen
  \bibfield  {author} {\bibinfo {author} {\bibfnamefont {F.}~\bibnamefont
  {Iga}}, \bibinfo {author} {\bibfnamefont {M.}~\bibnamefont {Kasaya}},
  \bibinfo {author} {\bibfnamefont {H.}~\bibnamefont {Suzuki}}, \bibinfo
  {author} {\bibfnamefont {Y.}~\bibnamefont {Okayama}}, \bibinfo {author}
  {\bibfnamefont {H.}~\bibnamefont {Takahashi}}, \ and\ \bibinfo {author}
  {\bibfnamefont {N.}~\bibnamefont {Mori}},\ }\href {\doibase
  http://dx.doi.org/10.1016/0921-4526(93)90591-S} {\bibfield  {journal}
  {\bibinfo  {journal} {Physica B:}\ }\textbf {\bibinfo {volume} {186}},\
  \bibinfo {pages} {419 } (\bibinfo {year} {1993})}\BibitemShut {NoStop}%
\bibitem [{\citenamefont {Takabatake}\ \emph {et~al.}(1990)\citenamefont
  {Takabatake}, \citenamefont {Teshima}, \citenamefont {Fujii}, \citenamefont
  {Nishigori}, \citenamefont {Suzuki}, \citenamefont {Fujita}, \citenamefont
  {Yamaguchi}, \citenamefont {Sakurai},\ and\ \citenamefont
  {Jaccard}}]{PhysRevB.41.9607}%
  \BibitemOpen
  \bibfield  {author} {\bibinfo {author} {\bibfnamefont {T.}~\bibnamefont
  {Takabatake}}, \bibinfo {author} {\bibfnamefont {F.}~\bibnamefont {Teshima}},
  \bibinfo {author} {\bibfnamefont {H.}~\bibnamefont {Fujii}}, \bibinfo
  {author} {\bibfnamefont {S.}~\bibnamefont {Nishigori}}, \bibinfo {author}
  {\bibfnamefont {T.}~\bibnamefont {Suzuki}}, \bibinfo {author} {\bibfnamefont
  {T.}~\bibnamefont {Fujita}}, \bibinfo {author} {\bibfnamefont
  {Y.}~\bibnamefont {Yamaguchi}}, \bibinfo {author} {\bibfnamefont
  {J.}~\bibnamefont {Sakurai}}, \ and\ \bibinfo {author} {\bibfnamefont
  {D.}~\bibnamefont {Jaccard}},\ }\href {\doibase 10.1103/PhysRevB.41.9607}
  {\bibfield  {journal} {\bibinfo  {journal} {Phys. Rev. B}\ }\textbf {\bibinfo
  {volume} {41}},\ \bibinfo {pages} {9607} (\bibinfo {year}
  {1990})}\BibitemShut {NoStop}%
\bibitem [{\citenamefont {Mason}\ \emph {et~al.}(1992)\citenamefont {Mason},
  \citenamefont {Aeppli}, \citenamefont {Ramirez}, \citenamefont {Clausen},
  \citenamefont {Broholm}, \citenamefont {St\"ucheli}, \citenamefont {Bucher},\
  and\ \citenamefont {Palstra}}]{PhysRevLett.69.490}%
  \BibitemOpen
  \bibfield  {author} {\bibinfo {author} {\bibfnamefont {T.~E.}\ \bibnamefont
  {Mason}}, \bibinfo {author} {\bibfnamefont {G.}~\bibnamefont {Aeppli}},
  \bibinfo {author} {\bibfnamefont {A.~P.}\ \bibnamefont {Ramirez}}, \bibinfo
  {author} {\bibfnamefont {K.~N.}\ \bibnamefont {Clausen}}, \bibinfo {author}
  {\bibfnamefont {C.}~\bibnamefont {Broholm}}, \bibinfo {author} {\bibfnamefont
  {N.}~\bibnamefont {St\"ucheli}}, \bibinfo {author} {\bibfnamefont
  {E.}~\bibnamefont {Bucher}}, \ and\ \bibinfo {author} {\bibfnamefont
  {T.~T.~M.}\ \bibnamefont {Palstra}},\ }\href {\doibase
  10.1103/PhysRevLett.69.490} {\bibfield  {journal} {\bibinfo  {journal} {Phys.
  Rev. Lett.}\ }\textbf {\bibinfo {volume} {69}},\ \bibinfo {pages} {490}
  (\bibinfo {year} {1992})}\BibitemShut {NoStop}%
\bibitem [{\citenamefont {Kasaya}\ \emph
  {et~al.}(1991{\natexlab{b}})\citenamefont {Kasaya}, \citenamefont {Tani},
  \citenamefont {Suzuki}, \citenamefont {Ohoyama},\ and\ \citenamefont
  {Kohgi}}]{doi:10.1143/JPSJ.60.2542}%
  \BibitemOpen
  \bibfield  {author} {\bibinfo {author} {\bibfnamefont {M.}~\bibnamefont
  {Kasaya}}, \bibinfo {author} {\bibfnamefont {T.}~\bibnamefont {Tani}},
  \bibinfo {author} {\bibfnamefont {H.}~\bibnamefont {Suzuki}}, \bibinfo
  {author} {\bibfnamefont {K.}~\bibnamefont {Ohoyama}}, \ and\ \bibinfo
  {author} {\bibfnamefont {M.}~\bibnamefont {Kohgi}},\ }\href {\doibase
  10.1143/JPSJ.60.2542} {\bibfield  {journal} {\bibinfo  {journal} {Journal of
  the Physical Society of Japan}\ }\textbf {\bibinfo {volume} {60}},\ \bibinfo
  {pages} {2542} (\bibinfo {year} {1991}{\natexlab{b}})},\ \Eprint
  {http://arxiv.org/abs/http://dx.doi.org/10.1143/JPSJ.60.2542}
  {http://dx.doi.org/10.1143/JPSJ.60.2542} \BibitemShut {NoStop}%
\bibitem [{\citenamefont {Ikushima}\ \emph {et~al.}(1999)\citenamefont
  {Ikushima}, \citenamefont {Yasuoka}, \citenamefont {Uwatoko},\ and\
  \citenamefont {Isikawa}}]{PhysRevB.60.14537}%
  \BibitemOpen
  \bibfield  {author} {\bibinfo {author} {\bibfnamefont {K.}~\bibnamefont
  {Ikushima}}, \bibinfo {author} {\bibfnamefont {H.}~\bibnamefont {Yasuoka}},
  \bibinfo {author} {\bibfnamefont {Y.}~\bibnamefont {Uwatoko}}, \ and\
  \bibinfo {author} {\bibfnamefont {Y.}~\bibnamefont {Isikawa}},\ }\href
  {\doibase 10.1103/PhysRevB.60.14537} {\bibfield  {journal} {\bibinfo
  {journal} {Phys. Rev. B}\ }\textbf {\bibinfo {volume} {60}},\ \bibinfo
  {pages} {14537} (\bibinfo {year} {1999})}\BibitemShut {NoStop}%
\bibitem [{\citenamefont {Das}\ and\ \citenamefont
  {Sampathkumaran}(1992)}]{PhysRevB.46.4250}%
  \BibitemOpen
  \bibfield  {author} {\bibinfo {author} {\bibfnamefont {I.}~\bibnamefont
  {Das}}\ and\ \bibinfo {author} {\bibfnamefont {E.~V.}\ \bibnamefont
  {Sampathkumaran}},\ }\href {\doibase 10.1103/PhysRevB.46.4250} {\bibfield
  {journal} {\bibinfo  {journal} {Phys. Rev. B}\ }\textbf {\bibinfo {volume}
  {46}},\ \bibinfo {pages} {4250} (\bibinfo {year} {1992})}\BibitemShut
  {NoStop}%
\bibitem [{\citenamefont {Sengupta}\ \emph {et~al.}(2012)\citenamefont
  {Sengupta}, \citenamefont {Iyer}, \citenamefont {Ranganathan},\ and\
  \citenamefont {Sampathkumaran}}]{Sengupta_2012}%
  \BibitemOpen
  \bibfield  {author} {\bibinfo {author} {\bibfnamefont {K.}~\bibnamefont
  {Sengupta}}, \bibinfo {author} {\bibfnamefont {K.~K.}\ \bibnamefont {Iyer}},
  \bibinfo {author} {\bibfnamefont {R.}~\bibnamefont {Ranganathan}}, \ and\
  \bibinfo {author} {\bibfnamefont {E.~V.}\ \bibnamefont {Sampathkumaran}},\
  }\href {\doibase 10.1088/1742-6596/377/1/012029} {\bibfield  {journal}
  {\bibinfo  {journal} {Journal of Physics: Conference Series}\ }\textbf
  {\bibinfo {volume} {377}},\ \bibinfo {pages} {012029} (\bibinfo {year}
  {2012})}\BibitemShut {NoStop}%
\bibitem [{\citenamefont {Zhang}\ \emph {et~al.}(2018)\citenamefont {Zhang},
  \citenamefont {Zhang}, \citenamefont {Chen}, \citenamefont {Lv},
  \citenamefont {Zhao}, \citenamefont {feng Yang}, \citenamefont {Chen},\ and\
  \citenamefont {Sun}}]{Zhang_2018}%
  \BibitemOpen
  \bibfield  {author} {\bibinfo {author} {\bibfnamefont {J.}~\bibnamefont
  {Zhang}}, \bibinfo {author} {\bibfnamefont {S.}~\bibnamefont {Zhang}},
  \bibinfo {author} {\bibfnamefont {Z.}~\bibnamefont {Chen}}, \bibinfo {author}
  {\bibfnamefont {M.}~\bibnamefont {Lv}}, \bibinfo {author} {\bibfnamefont
  {H.}~\bibnamefont {Zhao}}, \bibinfo {author} {\bibfnamefont {Y.}~\bibnamefont
  {feng Yang}}, \bibinfo {author} {\bibfnamefont {G.}~\bibnamefont {Chen}}, \
  and\ \bibinfo {author} {\bibfnamefont {P.}~\bibnamefont {Sun}},\ }\href
  {\doibase 10.1088/1674-1056/27/9/097103} {\bibfield  {journal} {\bibinfo
  {journal} {Chinese Physics B}\ }\textbf {\bibinfo {volume} {27}},\ \bibinfo
  {pages} {097103} (\bibinfo {year} {2018})}\BibitemShut {NoStop}%
\bibitem [{Note5()}]{Note5}%
  \BibitemOpen
  \bibinfo {note} {For non-cubic compounds, such as CeNiSn and CeRu$_4$Sn$_6$,
  the lattice parameter $a$ might have to be replaced with relevant nearest
  neighbour distances.}\BibitemShut {Stop}%
\bibitem [{\citenamefont {Georges}\ \emph {et~al.}(1996)\citenamefont
  {Georges}, \citenamefont {Kotliar}, \citenamefont {Krauth},\ and\
  \citenamefont {Rozenberg}}]{bible}%
  \BibitemOpen
  \bibfield  {author} {\bibinfo {author} {\bibfnamefont {A.}~\bibnamefont
  {Georges}}, \bibinfo {author} {\bibfnamefont {G.}~\bibnamefont {Kotliar}},
  \bibinfo {author} {\bibfnamefont {W.}~\bibnamefont {Krauth}}, \ and\ \bibinfo
  {author} {\bibfnamefont {M.~J.}\ \bibnamefont {Rozenberg}},\ }\href {\doibase
  10.1103/RevModPhys.68.13} {\bibfield  {journal} {\bibinfo  {journal} {Rev.
  Mod. Phys.}\ }\textbf {\bibinfo {volume} {68}},\ \bibinfo {pages} {13}
  (\bibinfo {year} {1996})}\BibitemShut {NoStop}%
\bibitem [{\citenamefont {Coleman}(1983)}]{PhysRevB.28.5255}%
  \BibitemOpen
  \bibfield  {author} {\bibinfo {author} {\bibfnamefont {P.}~\bibnamefont
  {Coleman}},\ }\href {\doibase 10.1103/PhysRevB.28.5255} {\bibfield  {journal}
  {\bibinfo  {journal} {Phys. Rev. B}\ }\textbf {\bibinfo {volume} {28}},\
  \bibinfo {pages} {5255} (\bibinfo {year} {1983})}\BibitemShut {NoStop}%
\bibitem [{\citenamefont {Hewson}(1993)}]{hewson}%
  \BibitemOpen
  \bibfield  {author} {\bibinfo {author} {\bibfnamefont {A.~C.}\ \bibnamefont
  {Hewson}},\ }\href@noop {} {\emph {\bibinfo {title} {The Kondo Problem to
  Heavy Fermions}}}\ (\bibinfo  {publisher} {Cambridge University Press},\
  \bibinfo {year} {1993})\BibitemShut {NoStop}%
\bibitem [{\citenamefont {Tomczak}\ \emph {et~al.}(2017)\citenamefont
  {Tomczak}, \citenamefont {Liu}, \citenamefont {Toschi}, \citenamefont
  {Kresse},\ and\ \citenamefont {Held}}]{Tomczak2017review}%
  \BibitemOpen
  \bibfield  {author} {\bibinfo {author} {\bibfnamefont {J.~M.}\ \bibnamefont
  {Tomczak}}, \bibinfo {author} {\bibfnamefont {P.}~\bibnamefont {Liu}},
  \bibinfo {author} {\bibfnamefont {A.}~\bibnamefont {Toschi}}, \bibinfo
  {author} {\bibfnamefont {G.}~\bibnamefont {Kresse}}, \ and\ \bibinfo {author}
  {\bibfnamefont {K.}~\bibnamefont {Held}},\ }\href {\doibase
  10.1140/epjst/e2017-70053-1} {\bibfield  {journal} {\bibinfo  {journal} {The
  European Physical Journal Special Topics}\ }\textbf {\bibinfo {volume}
  {226}},\ \bibinfo {pages} {2565} (\bibinfo {year} {2017})}\BibitemShut
  {NoStop}%
\bibitem [{Note6()}]{Note6}%
  \BibitemOpen
  \bibinfo {note} {For the temperature dependence of $\Delta (\omega )$ in
  Ce$_3$Bi$_4$Pt$_3$, see Ref.~\protect \rev@citealpnum {jmt_CBP_arxiv}; for
  spectral properties of Ce$_3$Bi$_4$Pt$_3$\ and Ce$_3$Bi$_4$Pd$_3$, see
  Refs.~\protect \rev@citealpnum {jmt_CBP_arxiv,Cao2019}.}\BibitemShut {Stop}%
\bibitem [{\citenamefont {Sinjukow}\ and\ \citenamefont
  {Nolting}(2002)}]{PhysRevB.65.212303}%
  \BibitemOpen
  \bibfield  {author} {\bibinfo {author} {\bibfnamefont {P.}~\bibnamefont
  {Sinjukow}}\ and\ \bibinfo {author} {\bibfnamefont {W.}~\bibnamefont
  {Nolting}},\ }\href {\doibase 10.1103/PhysRevB.65.212303} {\bibfield
  {journal} {\bibinfo  {journal} {Phys. Rev. B}\ }\textbf {\bibinfo {volume}
  {65}},\ \bibinfo {pages} {212303} (\bibinfo {year} {2002})}\BibitemShut
  {NoStop}%
\bibitem [{\citenamefont {Temmerman}\ \emph {et~al.}(2007)\citenamefont
  {Temmerman}, \citenamefont {Svane}, \citenamefont {Petit}, \citenamefont
  {Lüders}, \citenamefont {Strange},\ and\ \citenamefont
  {Szotek}}]{doi:10.1080/01411590701228703}%
  \BibitemOpen
  \bibfield  {author} {\bibinfo {author} {\bibfnamefont {W.~M.}\ \bibnamefont
  {Temmerman}}, \bibinfo {author} {\bibfnamefont {A.}~\bibnamefont {Svane}},
  \bibinfo {author} {\bibfnamefont {L.}~\bibnamefont {Petit}}, \bibinfo
  {author} {\bibfnamefont {M.}~\bibnamefont {Lüders}}, \bibinfo {author}
  {\bibfnamefont {P.}~\bibnamefont {Strange}}, \ and\ \bibinfo {author}
  {\bibfnamefont {Z.}~\bibnamefont {Szotek}},\ }\href {\doibase
  10.1080/01411590701228703} {\bibfield  {journal} {\bibinfo  {journal} {Phase
  Transitions}\ }\textbf {\bibinfo {volume} {80}},\ \bibinfo {pages} {415}
  (\bibinfo {year} {2007})},\ \Eprint
  {http://arxiv.org/abs/http://dx.doi.org/10.1080/01411590701228703}
  {http://dx.doi.org/10.1080/01411590701228703} \BibitemShut {NoStop}%
\end{thebibliography}


%

\end{document}